\global\let\ifmypprint\iffalse
\def\mypprint{\global\let\ifmypprint\iftrue}
\global\let\iftorefs\iffalse
\def\torefs{\global\let\iftorefs\iftrue}
\global\let\dofloatfig\iffalse
\def\floatthefig{\let\dofloatfig\iftrue}
\mypprint
\ifmypprint
\documentstyle[draft,pra,eqsecnum,aps,cite]{revtex}
\floatthefig
\else

  \iftorefs
    \catcode`\@=11
    
    \def\figure{\let\@capwidth\columnwidth\@float{figure}}
    \let\endfigure\end@float
    \@namedef{figure*}{\let\@capwidth\textwidth\@dblfloat{figure}}
    \@namedef{endfigure*}{\end@dblfloat}
    \catcode`\@=12
     \floatthefig
  \fi
\fi

\input epsf.tex

\tighten
\begin{document}
\renewcommand\citeleft{[}
\renewcommand\citeright{]}
\renewcommand\citepunct{, }
\makeatletter \renewcommand\@biblabel[1]{(#1)} \makeatother
\newcommand{\rref}[1]{{Eq. \ref{#1}}}
\newcommand{\CITEME}{\cite}
\newcommand{\A}{{\mathcal{A}}}
\newcommand{\Wk}{{\cal W}_k}
\newcommand{\beq}{\begin{equation}}
\newcommand{\eeq}{\end{equation}}
\newcommand{\goto}{\rightarrow}
\newcommand{\eiot}{{\rm e}^{i\omega t}}
\newcommand{\lo}{\ell(\omega)}
\newcommand{\ba}{\begin{eqnarray}}
\newcommand{\ea}{\end{eqnarray}}
\newcommand{\oh}{{1\over 2}}
\newcommand{\nut}{{\tilde\nu}}
\newcommand{\n}{{\bf{\hat{n}}}}
\newcommand{\ta}{{\bf{\hat{t}}}}
\newcommand{\ex}{{\bf{\hat e}_x}}
\newcommand{\ey}{{\bf{\hat e}_y}}
\newcommand{\dt}{\partial_{\tau}}
\newcommand{\af}{^{\{a,f\}}}
\newcommand{\p}{\partial}
\newcommand{\ie}{{\rm i.e.,}}
\newcommand{\cf}{{\it cf.}}
\newcommand{\Afelg}{{``A"}}
\newcommand{\Aus}{{A}}
\newcommand{\PHI}{\W}
\newcommand{\dg}{\partial_{\tau}g}
\newcommand{\da}[1]{\partial_{\alpha}^{#1}}
\newcommand{\enut}{{``{\nut}"}}
\newcommand{\en}{\eta}
\newcommand{\bk}{{\bf k}}
\newcommand{\bv}{{\bf v}}
\newcommand{\br}{{\bf r}}
\newcommand{\brp}{{\bf r}^{\perp}}
\newcommand{\bff}{{\bf f}}
\newcommand{\bz}{{\bf 0}}
\newcommand{\bZ}{{\bf Z}}
\newcommand{\bE}{\hat{\bf e}}
\newcommand{\ps}{\partial_{\sigma}}
\newcommand{\rg}{\sqrt{g}}
\newcommand{\gr}{{1\over\sqrt{g}}}
\newcommand{\eg}{{\rm e.g.,}}
\newcommand{\W}{{\cal W}}
\newcommand{\ka}{k \alpha}
\newcommand{\bfbj}{}
\newcommand{\bit}{}
\twocolumn[\hsize\textwidth\columnwidth\hsize\csname @twocolumnfalse\endcsname

\title{
Trapping and Wiggling: Elastohydrodynamics of Driven Microfilaments}
\author{Chris H. Wiggins,${*}$ Daniel X. Riveline,$^{\#}$ Albrecht Ott,$^{\#}$
and Raymond E. Goldstein$^{\S}$}
\address{${*}$Department of Physics, Princeton University, Princeton, NJ 08544,
cwiggins@princeton.edu}
\address{$^{\#}$Institut Curie, Section de Physique et Chimie, 11 Rue Pierre et
Marie Curie, 75231 Paris Cedex 05 France}
\address{$^{\S}$Department of Physics and Program in Applied Mathematics,
University of Arizona, Tucson, AZ 85721}

\date{\today}
\maketitle
\begin{abstract}
We present a general theoretical analysis of semiflexible filaments
subject to viscous drag or point forcing. These are the relevant
forces in dynamic experiments designed to measure biopolymer bending moduli.
By analogy with the ``Stokes problems" in hydrodynamics
(fluid motion induced by that of a wall bounding a viscous fluid), we
consider the motion of a polymer one end of which is moved in
an impulsive or oscillatory way.  Analytical solutions for the
time-dependent shapes of such moving polymers are obtained within
an analysis applicable to small-amplitude deformations.
In the case of oscillatory driving,
particular attention is paid to a characteristic length determined by
the frequency of oscillation, the polymer persistence length,
and the viscous drag coefficient.
Experiments on actin filaments manipulated with
optical traps confirm the scaling law predicted by the analysis
and provide a new technique for measuring the elastic bending
modulus.
A re-analysis of several published experiments on microtubules
is also presented.

\end{abstract}
\vskip2pc]

\section{Introduction}
\label{intro}

The attempts of theoretical physicists to contribute
in some useful way to the study of biology has,
so far, been most successful in systems in which all forces and
motion can be modeled and mathematized explicitly, or
those governed by equilibrium statistical mechanics, for which
equipartition can be invoked.  One
specific example of this success is
the analysis of structural microfilaments,
essentially one-dimensional
mechanical objects with no moving parts. Despite
this unassuming mechanical description, these semiflexible
biopolymers are responsible for innumerable essential
functions at the molecular and cellular level.

Depending on the bending modulus
of the filament in question, experiments on semiflexible biopolymers
largely rely on either mechanical or statistical
techniques. Microtubules, with a
persistence length $\sim 5$ mm, are quite amenable
to micromanipulation or forcing via hydrostatic drag. Actin
and nucleic acids, with persistence lengths
near $15 \mu$ and $50$ nm, respectively, fall
in the realm of statistical mechanics (note that
we are here addressing the {\it bending} elasticity,
not the {\it stretching} elasticity, another
area of great excitement and successes \CITEME{michelle,cluzel,carlos}).

However, much of the analysis of
elasticity in this context seems to ignore a fact known to the
ancients who developed the field: boundaries
matter for finite objects.  While often in physics boundary effects
can be ignored, the fact
that biopolymers are finite has important and
unexpected consequences and must be included carefully
the analysis. These effects appear because
dynamics of the elastica\footnote{a line of particles for which the
resistance to bending is a couple proportional to the curvature \CITEME{Love}}
are governed by fourth-order spatial derivatives.  This fact
forces us to rethink what are the most appropriate
basis functions in which to model dynamics. While
this question may seem purely academic, we will
see that it is quite crucial in attempts to
develop accurate dynamic methods for the quantitative
study of semiflexible biopolymers

Specifically, we couple elasticity theory
and overdamped viscous hydrodynamics (as is appropriate in
the biological context) to explore
{\it elastohydrodynamics}.  While experiments
coupling elasticity and drag have been presented
before \cite{felg,tony}, treatments of dynamics,
rather than mechanical equilibrium,
have been less than complete. Although equations
with the
appropriate
units will be
sufficient to
determine
the scales of forces and velocities,
if we wish to extract numbers from the experiments
it is necessary to have a thorough analysis.
The slenderness of the filaments allows us greatly to simplify the
hydrodynamics and arrive at a local partial differential
equation of motion.
Moreover, we will find that coupling to hydrodynamics
allows us to extend the range of {\it mechanical}
experiments to much smaller bending
moduli. For example, whereas
measurements
of actin's rigidity so far have been via
fluctuation analyses invoking equipartition and thus statistical mechanics,
we present here an experimental method which does not rely on
nonzero
temperature. Further, the method allows
investigation into questions which have been
raised as to whether or not actin can even be
treated as a semiflexible polymer, or is in
fact scale-sensitive \cite{Kas} or dynamic
in its elasticity. Such a purely mechanical
treatment obviates possible
complications to statistical treatments like dimensionality \cite{Ott},
correlations among
sampled images, or self-avoidance.

It is our hope that this new experimental method, as well
as the general analytic techniques here outlined,
will
contribute to the current exciting and active dialogue
among physicists and biologists regarding
the nature and numbers behind biopolymer rigidity
as well as the effects
of varying proteins and environments on elastic moduli.
Further, these new methods and analyses should prove useful
in the study of other examples of dynamic elastic
filaments, such as
supercoiled fibers of {\it B. subtilis} \cite{neil}.
We intend this investigation to be the necessary starting
point for such exciting extensions.

\dofloatfig \begin{figure} \epsfxsize=2.8 truein
\centerline{\epsffile{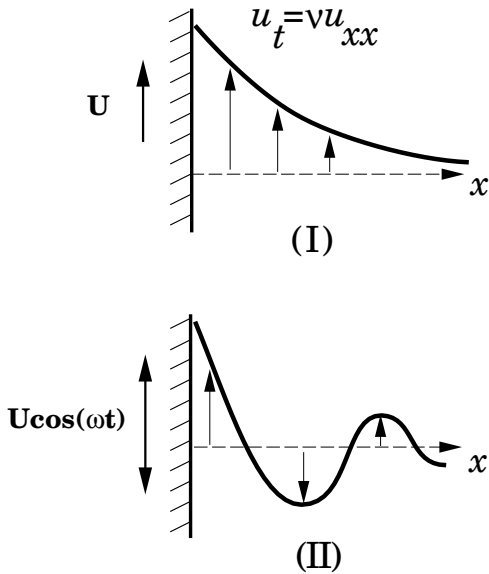}} \smallskip 
\caption[]{Geometry of Stokes problems I and II.} \label{fig_stokes} 
\end{figure} \fi

\label{stokes_intro}
A useful starting point for developing
the general mathematical treatment of the dynamics of an
elastic filament in a viscous medium will be
to consider the
simplest time dependencies possible.
To that end, recall the classic problems
introduced by G.G. Stokes in 1851 \cite{Stokes}, illustrated
in Fig. \ref{fig_stokes}, involving the motion of a viscous fluid
bounded by a wall that is either (I) moved impulsively or (II) oscillated.
These easily-solvable problems capture the essential ideas of
viscous diffusion of velocity.
The experimental geometry is such that the Navier-Stokes equation
for the velocity field $u(x,t)$ is simply the diffusion equation
\begin{equation}
u_t=\nu u_{xx}~,
\label{diff_eqn}
\end{equation}
where $\nu=\mu/\rho$ is the kinematic viscosity, and $\mu$ and
$\rho$ are the fluid viscosity and density. Subscripts on functions will
indicate differentiation throughout unless otherwise indicated.
The salient features of the solutions are the
relationships between length scales, time scales, and material
parameters. Specifically, in the impulsive case, the
velocity at any point $x$ and
time $t$ depends only on the ratio $x/(\nu t)^{1/2}$; likewise, in the
oscillatory case,
oscillations decay with a characteristic length
that scales as $\ell_S(\omega)=({\nu/\omega})^{1/2}$.

\begin{figure} \epsfxsize=3.0 truein
\centerline{\epsffile{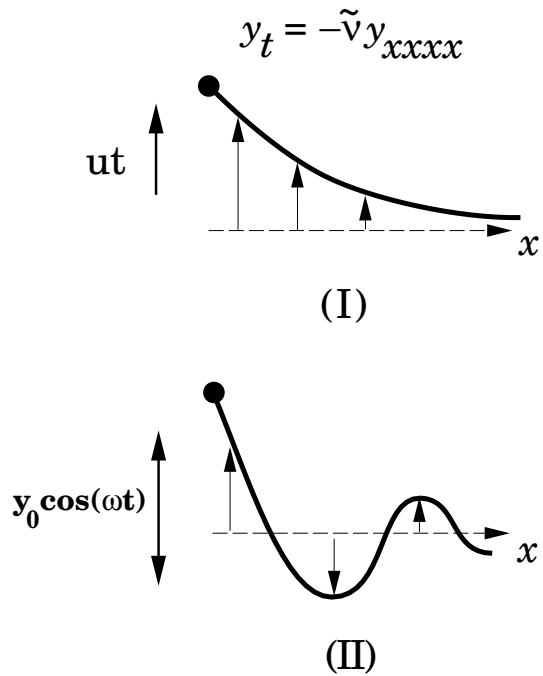}} \smallskip 
\caption[]{Geometry of elastohydrodynamic problems I and II.} 
\label{fig_polymer} \end{figure}

We introduce here the analogous two problems in elastohydrodynamics,
illustrated in Fig. \ref{fig_polymer}.  They involve (I) the deflection
of a polymer
anchored at one end following the instantaneous introduction of
a uniform fluid velocity $U$, and (II) the steady undulations of a polymer
one end of which is oscillated.  Rather than a diffusion equation as in the
Stokes problems, the dynamics of small deformations $y(x,t)$ of the filament
are governed by a {\it fourth-order} PDE of the form
\begin{equation}
y_t=-\tilde \nu y_{xxxx}~,\label{hyperdif}
\end{equation}
where $\tilde \nu=A/\zeta$ plays the role of a ``hyperdiffusion" coefficient,
with $A$ the bending modulus and $\zeta$ the drag coefficient.  This equation
has appeared before in the literature on semiflexible
biopolymers \cite{stanislas,Gittes}, primarily in the context of scaling
arguments for relaxation times; our goal here is to provide a complete solution
given arbitrary initial and boundary conditions as dictated by
experiment.\footnote{{\it Nota Bene:} in \cite{stanislas}, Eq. 2 should have a
minus
sign before the bending constant; as written, the equation is ill-posed.}

\dofloatfig  \fi

An analysis similar to that presented below
of the oscillatory passive elastica was carried out
a number of years ago by K. E. Machin \cite{mac1,mac2}, who considered the
motion
of a driven flagellum. Machin was interested specifically in
a semiinfinite
{\it active} flagellum which was bent with a set of boundary conditions
amenable
to analysis. Ours will be more malicious, but not subtle.

We will first derive the general equations of motion
for elastica embedded in viscous flow. We then apply this
dynamic to a number of scenarios.
Inspired by
``Stokes problems I and II" in fluid dynamics (SI and SII), we first solve
``problems I and II" of elastohydrodynamics (EHDI and EHDII), each
of whose dynamic mimics its hydrodynamic analogue. Problem I will
require a bit of mathematical construction to assist our physical
intuitions.
Specifically, we must construct a set of basis functions
appropriate to the equation of motion and specified boundary
conditions.  All the pleasant
features found when applying Fourier space
to unbounded or periodic systems are found here as well, in what we term
${\cal W}$-space.  Unlike Fourier space, ${\cal W}$-space respects both the
compact support and boundary conditions of the elastica and thus
diagonalizes the equation of motion.
We then discuss an experimental realization of problem II and its
analysis, which provides a new technique for the measurement of
a polymer's bending modulus.
Finally, we comment on some recent related experimental
work by a separate group, brought to our attention as the original
version of this paper was being completed.

\section{Elastic forces}

A bent elastic polymer exerts a restorative force
given by the functional derivative
of a bending energy
\beq
{\bf f}_{\cal E}=-{\delta{\cal E}\over\delta{\bf r}}
\label{f_ge},
\eeq
which for an
elastic curve
may be expressed as\footnote{Note that we may also include any forces of
constraint,
such as a Lagrangian tension to enforce inextensibility \cite{reg}, but
such terms will be of higher order in the curvature than
we will consider in this investigation.}
\beq
{\cal E}=\oh A\int_0^L\!\! ds \kappa^2~.
\label{eex}
\eeq
Here, $A$ is the bending stiffness constant, with units of
energy~$\times$~length. It may also be expressed as the product $EI$ of
the Young's modulus $E$ and the moment of inertia $I$\cite{feynman}.
For a polymer of
persistence length $L_p$ at absolute temperature $T$,
exploring all configurations in $D$ dimensions, we may also
derive by equipartition the equivalence $A=(D-1)k_BTL_p/2$.

\begin{figure} \epsfxsize=3.0 truein
\centerline{\epsffile{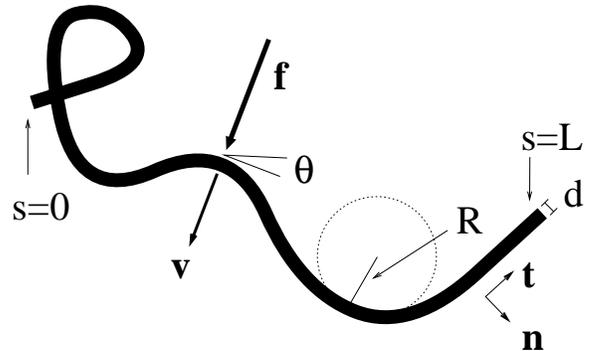}} \smallskip 
\caption[]{Geometry of an elastic filament: $R$=local radius of 
curvature; $\bf{t},\bf{n}$=unit tangent and normal, 
$\theta=\cos^{-1}(\ta\cdot\ex)$; $d$=diameter of the filament; 
arclength $s$ varies from $0$ to $L$, the total arclength.  
Within the approximations of slender-body hydrodynamics, 
a local anisotropic proportionality is satisfied between an 
external force per unit length $\bf f$ and the velocity $\bf v$.} 
\label{fig_geometry} \end{figure}

Henceforth we will consider elastic filaments that lie in a
plane. This is the geometry most well-suited to data-taking via microscopy.
 The curvature $\kappa$ may then be expressed exactly
as $d{\theta}/ ds$,
where $\theta$ is the angle between the tangent to the curve
and some fixed axis (see Fig. \ref{fig_geometry});
the $x$-axis is a convenient choice, for which
$\tan\theta=(dy/ds)/(dx/ds)=dy/dx$,
which implies
\beq
\kappa={y_{xx}\over\left(1+y_x^2\right)^{3/2}}~.
\eeq
Taking the functional derivative of the energy
(\ref{eex}), we derive the force and boundary conditions
\ba
{\bf f}_{\cal E}=A\left(\kappa_{ss}+\oh\kappa^3\right)\n~,\\
\kappa=\kappa_s=0~~{\rm at~free~ends}.\label{exact_bcs}
\ea
These boundary conditions indicate the torquelessness
and forcelessness, respectively,
at free ends of an elastica \cite{Wein}.
At hinged or clamped ends different boundary
conditions hold, as will be discussed below.

\dofloatfig  \fi

We may linearize these results for small deviations from a horizontal line,
($y_x\ll1$) or equivalently linearize Eq. \ref{eex} to find $\kappa\simeq
y_{xx}+{\cal O}\left(y_x^2\right)$ and
\beq
{\bf f}_{\cal E}\simeq-Ay_{xxxx}\ey+{\cal O}\left(y_x^2\right)~,
\eeq
(where $\ey$ is the unit vector in the
$y$ direction) with boundary conditions
\beq
y_{xx}=0~{\rm and}~y_{xxx}=0~~{\rm at~free~ends}\label{bclin}.
\eeq

The specification of the filament dynamics is complete upon definition
of the hydrodynamic drag which balances  ${\bf f}_{\cal E}$.
We now turn to this problem.

\section{Slender-body hydrodynamics}
\label{sbh}
We wish to consider experiments taking
place on cellular biological scales, with typical lengths $L$
in microns, times $t$ in seconds, and
dynamic viscosity $\mu$  that of water, in centipoise.
The nondimensional parameter
characterizing the
hydrodynamic behavior, the Reynolds number, is
\nobreak{$UL/\nu\sim L^2/t\nu\sim 10^{-8}/10^{-2}\sim 10^{-6}$.}
We are thus safely in the low-Reynolds number or `Stokesian'
regime. In this Aristotelian overdamped limit, forces balance velocities
rather than accelerations. For a body whose length is much
greater than its width, the well-developed
set of calculations known as `slender-body
hydrodynamics' applies \cite{KnR,cox1,cox2}.
If this filamentous polymer has diameter $d$,
length $L$, and an aspect ratio $
d/L\ll 1$, we have the simplified, local, anisotropic
proportionality
between the drag force ${\bf f}_d$ and the velocity ${\bf r}_t$,
\beq
{\bf f}_d=
\zeta\left[\n\n+\beta\ta\ta\right]\cdot
\left({\bf r}_t-{\bf u}\right).
\label{eqmot}
\eeq
Here, ${\bf f}_d\left(s\right)$ is a force {\it per unit length} exerted
on the filament, and
$\n\left(s\right)$ and $\ta\left(s\right)$ are unit vectors in the normal and
tangential
directions at arclength $s$ along the polymer.
The velocity of the polymer is denoted ${\bf r}_t\left(s\right)$,
and ${\bf u}$ is any background velocity which may be present in
the problem; the drag should be a function of the former
relative to the latter.
The anisotropy, evident when dragging a pencil through molasses,
between motions parallel and perpendicular to a slender object's long axis
is embodied by
the parameter $\beta$, which depends logarithmically on the
aspect ratio, with asymptotic behavior
\beq
\lim_{d/L\rightarrow0}\beta=\oh~.\label{betadisc}
\eeq
An active field of applied mathematics in the last twenty years has
been calculations of the viscous drag coefficient
$\zeta$ for various slender shapes \cite{KnR,cox1,cox2,lh,childress,mikented}.
The limiting behavior for small $d/L$ has the form
\beq
\zeta={4\pi\mu\over{\ln\left(L/d\right)+c}}~.\label{dragcoeff}
\eeq
where $c$ is a constant of order unity which depends on the shape
of the body.

Now that we have defined the hydrodynamic drag, we
demand that it balances
elastic forcing (${\bf f}_d={\bf f}_{\cal E}$)
to derive the equation of motion:
\beq
\zeta\left[\n\n+\beta\ta\ta\right]\cdot
\left({\bf r}_t-{\bf u}\right)
=
A\left(\kappa_{ss}+\oh\kappa^3\right)\n.
\label{fund_eq_mot}
\eeq
Linearizing the expression for the drag (\ref{eqmot}) for nearly-straight
polymers ($y_x\ll 1$), and noting that its tangential components
are of order $y_x^2$, the dynamic reduces to
\beq
\zeta\left(y_t-u\right)=-Ay_{xxxx}~.\label{EqMotInhom}
\eeq
Here, $u\equiv  {\bf u}\cdot\ey$.
In the absence of any background flow,
\beq
y_t= -\nut y_{xxxx},\label{eqmotdim}
\eeq
where $\nut=A/\zeta$ as in Eq. \ref{hyperdif}.
This is the simplest linearized expression of elastohydrodynamics:
elastic forces, characterized by a fourth spatial derivative,
balance viscous drag. It shares many similarities with the diffusion
equation (\ref{diff_eqn}), and may be thought of as ``hyperdiffusion"
of displacement in analogy with hyperviscosity.

\section{Elastohydrodynamic problem I}
\label{spI_II}

Now that we have established the equation of motion appropriate
to these elastohydrodynamic analogs,
we
recall the solutions to the fluid dynamics problems SI and SII
in hopes of exploiting the analogy as much as possible.
In Stokes I (SI), a semiinfinite
plane of fluid is forced by a wall which is motionless for time
$t<0$ and has velocity $U \ey$ for $t>0$. In
Stokes II (SII), the wall oscillates as $U\cos\left(\omega t\right)\ey$, and
we solve for the behavior after transients have died away.

As illustrated in Fig. \ref{fig_stokes}, velocity gradients are in the
$x$-direction in both cases, and
hence perpendicular to the direction of flow (along the $y$-axis).
In the absence of an imposed pressure gradient, the Navier-Stokes equation for
the fluid velocity $u\left(x,t\right)$ parallel to the wall
is simply the {\it linear} diffusion equation (\ref{diff_eqn}),
$u_t=\nu u_{xx}$.

A convenient method of solving SI with the associated boundary condition
is to postulate a scaling solution inspired by dimensional analysis:
$u\left(x,t\right)=UF\left(\xi\right)$, with $\xi=x/\left(\nu t\right)^{1/2}$.
The scaling function $F$
then obeys a nonautonomous ordinary differential equation
\begin{equation}
-{1\over 2}\xi F_{\xi}=F_{\xi\xi}~.
\label{ode_stokesI}
\end{equation}
The solution to this is
$F={\rm erfc}\left(\xi/\sqrt{2}\right)$,
where ${\rm erfc}$ is the complementary error function. The dedicated
may arrive at the same result more laboriously via Laplace transforming
\rref{diff_eqn}.

Armed with some understanding of SI, we now
turn to  problem I of elastohydrodynamics (EHDI).
In problem I, we  consider an elastic filamentous polymer
which is anchored at the origin.
For $t<0$ it lies along
the line segment $\{y=0; \ 0<x<L\}$.
We then may consider forcing the filament by moving
one end relative to the fluid (moving the anchor) or moving
the fluid relative to the polymer (moving, for example, the cover slip).
We will first attempt to do this in a way as analogous to SI as possible.

\subsection{Semiinfinite polymer}

Consider first moving the anchor relative to the fluid with
velocity ${\bf u}=u\ey$. This may
be thought of as moving a bead optically trapped to one end,
or some other
micromanipulation.
We now observe that the equation of motion, $y_t=-
{\tilde\nu}y_{xxxx}$, inspires a scaling solution of the form
$y=utF\left(\xi\right)$, where $\xi\equiv  x/\left(\nut t\right)^{1/4}$.
Inserting this
special form of the solution, the equation becomes
\beq
F-{1\over 4}\xi F_{\xi}=-F_{\xi\xi\xi\xi},\label{ode_EHDI}
\eeq
an equation not unlike \rref{ode_stokesI}.

The solution $F$ must admit the conditions that it and
its derivatives approach
$0$ for a fixed time as $x\goto\infty$, so
\beq
\lim_{\xi\goto\infty}
\{F\left(\xi\right),
F_{\xi}\left(\xi\right),
F_{\xi\xi}\left(\xi\right),
\ldots
\}
=0~.
\eeq
Moreover, specifying the anchor position implies
\beq
ut=y\left(0,t\right)=utF\left(0\right)~~\Rightarrow~~F\left(0\right)=1~.
\eeq
Depending on whether we wish to describe a filament
anchored by, for example, an axoneme or a bead, we demand $y_x\left(0\right)=0$
or $y_{xx}\left(0\right)=0$,
implying $F_{\xi}(0)=0$ or $F_{\xi\xi}\left(0\right)=0$, respectively.

\dofloatfig \begin{figure} \epsfxsize=3.0 truein
\centerline{\epsffile{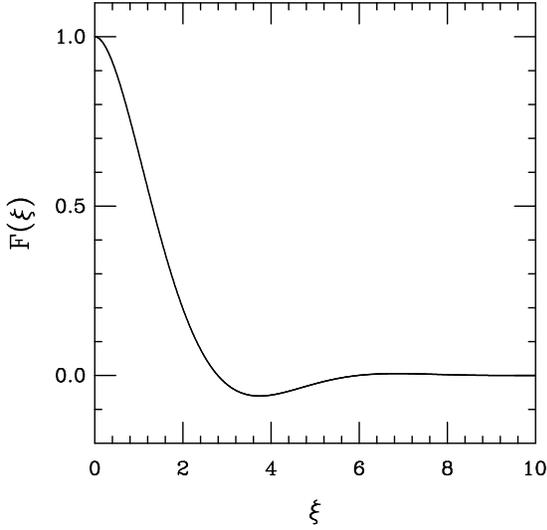}} \smallskip 
\caption[]{Scaling solution to (\ref{diff_eqn}) for a semiinfinite 
clamped polymer. The meaning of ``scaling solution'' is that $y$, 
$x$, and $t$, are suitably rescaled such that, if a friend were to 
imprint this image on silly putty, and run towards you with a 
velocity $\sim t^{-5/4}$ while stretching the putty in the $y$ direction 
a length $\sim t^{3/4}$, you would see the predicted shape of the 
semiinfinite elastica. For the most accurate results, have your friend 
start infinitely far away. This complication is obviated by considering 
the finite elastica.} \label{scal_soln} \end{figure} \fi

Note that, had we instead chosen to consider a fixed {\it anchor}
and a moving {\it fluid},
we would merely implement two changes. First, the equation of motion becomes
$\left(y_t-u\right)= -{\tilde\nu}y_{xxxx}$. Second, the boundary condition
becomes
$y\left(0\right)=0$. However, introducing a Galilean change
of frame $y'\left(x,t\right)=y-ut$,
we see that we recover the equivalent equation of motion, boundary conditions,
and thus problem. This
relationship
between boundary conditions and equations
of motion will allow us to apply one analysis to a number of experiments.
We will use the terms ``homogeneous" and ``inhomogeneous" referring to
both boundary conditions and equations of motions since we may transform
from one to the other so simply.

Though nonautonomous, Eq. (\ref{ode_EHDI}) is linear in $F$ and
quite amenable to numerical solution. A plot of the solution for
the clamped ($y_x(0)=0$) polymer
is shown in Fig. \ref{scal_soln}. Clearly, the function
satisfies our qualitative intuitions about the shape: an initially
straight curve slopes towards $0$ as the argument grows,
where all derivatives vanish. The appearance of the wiggle
leads us more carefully to consider the realm of validity of our solution.
Given our scaling ansatz $y=utF(\xi)$,
we see that the slope $y_x=utF_{\xi}/(\nut t)^{1/4}$ grows
in time without bound, thus failing to meet the criterion
on which the linearization of \rref{fund_eq_mot}
was predicated: $y_x\ll 1$.
Intuitively, we
suspect that the neglected nonlinear terms would suppress these
undulations.

Noting that $|F_{\xi}|$ attains its largest value of
$\simeq 0.51$ for $\xi^*\simeq 1.1$, we see that we can only
trust the solution to initial times such that
\beq
t\ll{\nut^{1/3}\over \left(uF_{\xi}\left(\xi^*\right)\right)^{4/3}}
\simeq 2.5{\left({\nut/ u}\right)^{1/3}\over u}
\eeq
For an experimentally convenient velocity of $\sim 10 \mu/s$,
this time is order a fraction of a second for actin, and
a second for microtubules. We have thus seen that the
obvious extension of the hydrodynamic analogue
is not necessarily the correct
approach -- a conclusion which will be reached repeatedly
in this investigation.

Enjoyable though it is to contemplate the long-time behavior of the
semiinfinite dragged elastica,
a more profitable inquiry would be the experimentally
realizable case of a finite polymer. We therefore turn now to the
case of the finite elastica, clamped at one end and free at the
other, and subject to impulsive hydrodynamic drag.
Our analysis is applicable to
experiments in
which either the cover slip is moved or, as a special case,
in which the elastica is allowed
to relax from some initial conditions in the absence of flow.

\subsection{Finite polymer}
\label{EHDIsec}
In order to make the mathematics as transparent as possible, we
first nondimensionalize the equation of motion (\ref{EqMotInhom}).
Distances in $x$ are rescaled by the total length $L$, time by
the elastohydrodynamic time scale $\zeta L^4/A$, and distances
in $y$ by the velocity times this time scale:
\beq
x=\alpha L~, \ \ \  t=\tau{\zeta L^4\over A}, \ \ \
y\left(x,t\right)=u{\zeta L^4\over A}h\left(\alpha,\tau\right)~.
\eeq
The governing equation, $(y_t-u)=-\nut y_{xxxx}$, then becomes
\beq
(h_{\tau}-1)=-h_{\alpha\alpha\alpha\alpha}~.
\label{ndim}
\eeq
Like any other inhomogeneous differential equation, the general solution
of (\ref{ndim}) must
be the homogeneous solution
($h_{\tau}=-h_{{\alpha\alpha\alpha\alpha}}$)
plus the particular solution
($h_{\tau}-1=-h_{{\alpha\alpha\alpha\alpha}}$).
We indicate this by defining
\beq
h=f+g
\eeq
where $g$ and $f$ satisfy
\beq
{g}_{\tau}=-g_{{\alpha\alpha\alpha\alpha}},~~~
{f}_{\tau}-1=-f_{{\alpha\alpha\alpha\alpha}}\label{feqn}.
\eeq
We will find that these two functions have markedly different
qualitative behavior.

\subsubsection{Homogeneous solution}

The homogeneous equation is
$
{g}_{\tau}=-g_{{\alpha\alpha\alpha\alpha}}
$.
Well-versed in the litany of Fourier transforms,
we first left-multiply by an as-yet arbitrary function
$\PHI_k\left(\alpha\right)$ (where $k$ indicates a parameter
rather than a derivative)
and integrate over the domain of $\alpha$,
\beq
\int_0^1\! d \alpha\PHI_k{\dg}=
-\int_0^1\!d\alpha\PHI_k\da{4}g~.
\eeq
Integrating by parts we obtain
\ba
\partial_{\tau}\int_0^1\!d\alpha\PHI_kg&
=-& \int_0^1d\alpha\da{4}(\PHI_k)g\nonumber\\
&-&\PHI_k\da{3}g\bigl|_0^1+\da{1}\PHI_k\da{2}g\bigl|_0^1\nonumber\\
&-&\da{2}\PHI_k\da{1}g\bigl|_0^1
+\da{3}\PHI_kg\bigl|_0^1.
\label{mess}
\ea
Observe that the integration by parts of the fourth-order derivative
introduces $8$ separate surface terms.
The boundary
conditions implied by the functional derivative (\ref{exact_bcs}) dictate
the vanishing of the second and third derivatives
at the free end ($x=L$).
Requiring $g$ to satisfy this condition eliminates two of the $8$
terms.

The left end of the polymer is clamped at the origin, so
$y\left(x=0\right)=y_x\left(x=0\right)=0$.
Demanding this behavior of $g$ eliminates two additional surface
terms. We now choose $\PHI_k$
to satisfy the same boundary conditions
as $y$ and $g$, and therefore $h$:
\ba
\PHI_k\left(0\right)=0&,~~\partial_{\alpha}\PHI_k\left(0\right)=0,&\nonumber\\
\partial^2_{\alpha}\PHI_k\left(1\right)=0&,
{}~\partial^3_{\alpha}\PHI_k\left(1\right)=0.&\label{bcs}
\ea
This annihilates
the remaining 4 surface terms in (\ref{mess}). Finally,
we choose $\PHI_k$ to obey
\beq
\da{4}\PHI_k=k^4\PHI_k~.
\label{phik_difeqn}
\eeq
Defining
$g_k\equiv \int_0^1\! d\alpha\PHI_kg~,
$
the equation of motion
becomes
${\dg_k}= -k^4g_k,$
the solution to which is
\beq
g_k\left(\tau\right)=g_k\left(0\right){\rm e}^{-k^4\tau}~.\label{gsol}
\eeq
If we wish to describe the dynamics in such terms, we must construct
the $\PHI_k$, which necessitates that we identify the allowed values of
$k$. We turn now to this problem.

\subsubsection{Explicit construction of the $\PHI_k$}
A moment's thought reveals that the $\PHI_k$ can not
be constructed out of simply the familiar sin's and cos's of
Fourier
space,
which are incompatible with boundary
conditions in which successive derivatives vanish (\cf~ Eq. \ref{exact_bcs}).
A countably
infinite family of such $\PHI_k$ can, however, be constructed by
including
hyperbolic functions as well
in the basis of the function space.
The general solution of
\rref{phik_difeqn} is
\ba
\PHI_k=a_1&\sin\left(k\alpha\right)&+a_2\cos\left(k\alpha\right)\nonumber \\
+a_3&\sinh\left(k\alpha\right)&+a_4\cosh\left(k\alpha\right).\label{GenSoln}
\ea
The expression has four unknowns, as a solution to a fourth-order
problem must. The $\alpha=0$
(left end)
boundary conditions yield
$
a_4=-a_2$ and $a_3=-a_1.
$
The last two conditions,
$\da{2}\PHI_k\left(1\right)=0$ and $\da{3}\PHI_k\left(1\right)=0$,
can be expressed as a matrix equation
\beq
\pmatrix{\sinh k + \sin k&
\cosh k  + \cos k \cr
\cosh k + \cos k &
\sinh k  - \sin k  }
\pmatrix{a_1\cr a_2}=0.
\label{mat}
\eeq
Since the matrix has a zero eigenvalue, it must be singular.
This solvability condition is
\beq
\cos k = -{1\over \cosh k}~. \label{solv_cond}
\eeq

 \dofloatfig \begin{figure} \epsfxsize=3.3 truein
\centerline{\epsffile{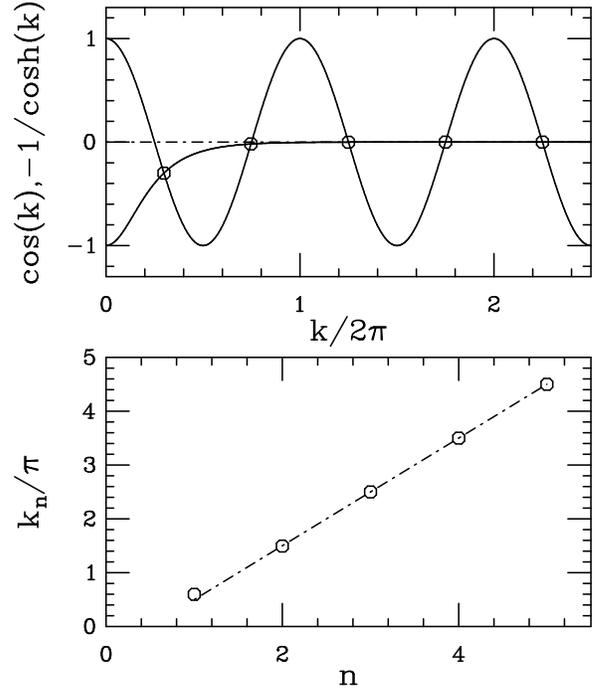}} \smallskip 
\caption[]{Solution of the transcendental equation for the roots $k_n$. 
(a) Graphical construction of the solution.  (b) The roots (circles) 
asymptote to $k_n=\pi\left(2n-1\right)/2$ (dashed line).} \label{solv_fig} 
\end{figure} \fi
As Fig. \ref{solv_fig} illustrates,
this transcendental equation has an infinite number of
solutions. For large values of $k$, as
$1/\cosh k \rightarrow 0$, the solutions approach
the solutions of the Fourier-like solvability condition
$\cos k=0$, \ie~ $k_{n+1}\goto{\pi\over 2}+\pi n$.
The first few solutions are
\ba
k_1&\simeq {\pi\over 2}+.304\simeq 1.875~,
&k_2\simeq {3\pi\over 2}-.018\simeq 4.694~,\nonumber\\
k_3&\simeq {5\pi\over 2}+.001\simeq 7.855,~
&k_4\simeq{7\pi\over 2}\simeq  10.996~, \ldots
\ea

\begin{figure} \epsfxsize=3.3 truein
\centerline{\epsffile{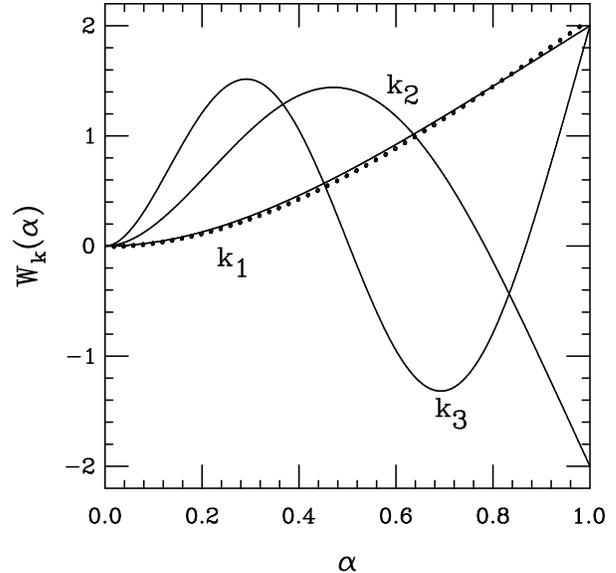}} \smallskip 
\caption[]{The first three eigenfunctions for EHD problem I. 
The dotted line indicates the normalized third-order polynomial 
describing an elastica bent by a point force at the right end. 
Note the surprising overlap with $\PHI_{k_1}$, as will be exploited 
in section \ref{initdata}.} \label{fte} \end{figure}

We can insert either row of the LHS of (\ref{mat}) to solve for the
relationship between $a_2$ and $a_1$, which we will
express here as
\beq
a_2=-\left({\sin k +\sinh k \over \cos k + \cosh k}\right) a_1.
\eeq
The constants themselves are functions of $k$.
Note also that as $k \rightarrow \pm\infty$, $a_2\rightarrow\pm 1$.
We now choose $a_1$
such that $\PHI_k$ is normalized, {\ie}
$\int_0^1d\alpha\PHI_k^2=1$.
The first three eigenfunctions are shown in Fig. \ref{fte}.

We have now completely and
explicitly constructed the family of functions $\PHI_k$
which are normalized eigenfunctions of $\p_{\alpha}^4$ with particular
boundary conditions.
Note that had we chosen some other set of boundary conditions,
a different solvability condition and eigenfamily would
have resulted ({\cf} Appendix \ref{family}).
For example, in the case of the elastica with free ends, we employ an
expansion of $y$ with the basis functions of Eq. (\ref{A5}).  Note that if we
differentiate this expression and define $\alpha_n=k_n/2$, then we
recover Eq. 28 of \cite{Gittes}.  The expansion in
terms of $k$ has the advantage of a single solvability condition
$\cos(k_n)\cosh(k_n)=1$,
rather than the two conditions $\tanh(\alpha_n)=(-1)^n\tan(\alpha_n)$.
The latter two conditions
may be derived from the former via half-angle formulae.

This operator is self-adjoint ({\cf} Appendix \ref{prop_app})
and thus the eigenfunctions
constitute a complete basis in function space onto which we may
project initial
data and relate to later-time solutions
via \rref{gsol} in
the standard Green's function way:
\ba
g\left(\alpha,\tau\right)&=&\sum_k
|\PHI_k\rangle\langle\PHI_k|g\left(\alpha,\tau\right)\rangle \nonumber\\
&=&\sum_k|\PHI_k\rangle g_k\left(0\right){\rm e}^{-k^4\tau}\nonumber\\
&=&\int_0^1\! d\alpha'{\cal G}\left(\alpha,\alpha';\tau\right)
g\left(\alpha',0\right)~,
\ea
where the Green's function is
\beq
{\cal G}(\alpha,\alpha';\tau)
=\sum_k
\PHI_k\left(\alpha\right)
\PHI_k\left(\alpha'\right)
{\rm e}^{-k^4\tau}
{}.
\label{exsol}
\eeq
This is the exact solution of the linearized
homogeneous equation.
Note that the compact support and the boundary conditions
break translation invariance, reflected in the fact ${\cal G}$
can not be expressed as ${\cal G}\left(\alpha-\alpha';\tau\right)$.

\dofloatfig  \fi

We note from the exact solution (\ref{exsol}) that each mode $g_k$
decays independently and exponentially with time. This is to be compared
with diffusive
problems, in which each mode
decays exponentially in time, {\it except} for the zero (average) mode,
which is constant.
In this experiment, the boundary conditions
are incompatible with the existence of a zero mode.
The system ``hyperdiffuses"
to homogeneity.

\subsubsection{Particular solution}

Since $g(\alpha,\tau)$ decays to zero as $\tau\to\infty$,
we have
\beq
\lim_{\tau\goto\infty}h\left(\alpha,\tau\right)=
f\left(\alpha,\infty\right).
\eeq
Thus the particular solution, if it exists, must be the long-time solution;
no matter what initial data, the solution to the inhomogeneous
equation of motion (\ref{ndim}) must be the global attractor.

By the same argument,
we must only construct a solution for it to be the correct one.
We first look for a time-independent function, for which
Eq. \ref{feqn}
becomes
$
1=f_{\alpha\alpha\alpha\alpha}$.
The unique solution consistent with the boundary conditions (\ref{bcs}) is
\beq
f\left(\alpha\right)={1\over24}\left(\alpha^4-4\alpha^3+6\alpha^2\right)\equiv
{\bar{h}}\left(\alpha\right)
\eeq

\subsubsection{Digression on boundary conditions}
It is useful
to
pause
momentarily
and reflect on the importance of
boundary conditions and how they will enter into the analysis. To this
end, we consider
an analogous experiment in which the left side of the polymer
is hinged rather than clamped.
This would be realized by holding the polymer
in an optical trap, for example, rather than some torque-exerting
anchor like an axoneme fixed to a cover slip. For this experimental setup
we replace the boundary condition
$h_{\alpha}\left(0,\tau\right)=0$
with
$h_{\alpha\alpha}\left(0,\tau\right)=0$.

Intuitively, we expect the polymer to align itself
with the flow as $t\goto\infty$ in the absence of any
torque at the left end. Mathematically, we may think of
the change in
boundary conditions via some curious pathologies.
The first complication is that
there does not exist a time independent fourth-order
polynomial in $\alpha$
which is consistent with the boundary conditions. This
prevents us from constructing a static attractor for the problem.
However, we note that the
equation of motion is solved by a fifth-order, {\it time-dependent}
polynomial in $\alpha$:
\beq
{\bar h}\left(\alpha,\tau\right)={3\over 2}\left(\tau+C\right)\alpha+{1\over
4!}
\left(-\alpha^3+\alpha^4-{3\over 10}\alpha^5\right).\label{EHDIP}
\eeq
To derive this poynomial, we first express the solution in the
form $\sum_n c_n\left(\tau\right)\alpha^n$.
However, we are considering a Stokesian dynamic, in which
the time dependence is first-order and the equation of motion
is linear; the driving is a constant in time
and thus time should enter only linearly into any steady
state solution for the position ({\cf} \cite{LnL}).
Constraining
$
\partial_t^2 c_n=0
$
and respecting the relationship between
$\p_{\tau}c_n$ and $c_{n+4}$
dictated by \rref{ndim},
we uniquely
specify the polynomial up to the constant $C$.

The fact that our long-time solution has an arbitrary constant
should lead us to rethink splitting the solution into a polynomial
and a set of only-decaying modes. Returning to the set of eigenvalues
of $\partial_{\alpha}^4$, we discover a second complication for
this new boundary condition: there now exists a zero mode -- a
nontrivial solution,
consistent with the boundary conditions, to the
equation $\partial_{\alpha}^4\PHI_0=0$, \ie~ the normalized
polynomial $\sqrt{3}\alpha$.
We then may choose $C$ to eliminate the overlap
of $\PHI_0$ with the initial data, \ie~
$0=g_0
=\langle g|\PHI_0\rangle=
\langle h(\alpha,0)|\PHI_0\rangle-\langle {\bar h}(\alpha)|\PHI_0\rangle$.

We now see how the change in boundary conditions creates drastically
different physical behavior. The long-time solution contains a term describing
a straight line whose slope grows with velocity $3u/2L$ without bound.

In order further to illustrate the relationship between a change in
boundary conditions and the qualitative behavior,
note that the dynamics, even in the presence of an
inhomogeneous equation of motion, can be cast in terms of the functional
derivative of an energy:
\ba
{h}_{\tau}&=&-{\delta {\cal F}\over\delta h\left(\alpha\right)},\nonumber \\
{\cal F}&=&
\int_0^1\! d\alpha'\left [-h+\oh \left(h_{\alpha\alpha}\right)^2\right].
\ea
The first term represents the drag force acting in the
positive $y$ direction, while the second is
simply the nondimensionalization of the
elastic bending energy term from which we originally derived
the equation of motion. We then find
\beq
{{\cal F}_{\tau}}
=\int_0^1 d\alpha
	{h_{\tau}}
	{\delta{\cal F}\over\delta h\left(\alpha\right)}
=-\int_0^1 d\alpha \left(h_{\tau}\right)^2,
\eeq
indicating ${\cal F}$ is a monotonically decreasing function in
time \cite{CnH,reg
}.

We must
now
only show
that for the clamped (hinged)
polymer this energy functional
is (is not) bounded from below. Rewriting
$
h=\sum c_n(\tau)\alpha ^n,
$
we can evaluate the energy explicitly as
\beq
{\cal F}=-\sum_{n=0}^\infty{c_n\over n+1}+{\cal R}~,
\label{energy}
\eeq
where
\beq
{\cal R}=\oh\sum_{m,n=2}^\infty {n\left(n-1\right)m\left(m-1\right)\over
n+m-3}c_mc_n~.
\label{R_define}
\eeq
If $h\left(0\right)=h_x(0)=0$, $c_0=c_1=0$. \rref{energy} then becomes
\beq
{\cal F}
=-\sum_{n=2}^\infty{c_n\over n+1}+{\cal R}~.
\eeq
We see that we can not simply make the first term arbitrarily negative
by introducing a large and positive $c_j$ for some $j$, since this
term will appear quadratically (and always positively) in the second
term. However,
replacing the condition $h_{\alpha}\left(0\right)=0$
with $h_{\alpha\alpha}\left(0\right)=0$ changes
the condition $c_1=0$ to $c_2=0$, and the energy becomes
\beq
{\cal F}
=-{c_1\over 2}-\sum_{n=3}^\infty{c_n\over n+1}+{\cal R}~.
\eeq
Now the energy can become arbitrarily negative
if $c_1=h'(0)$ becomes arbitrarily positive.
A divergent slope simply means the curve points straight up, in
accord with our intuition for the long time behavior of a polymer free to
rotate
in some background flow.
Note such a long-time behavior means
leaving the small-$h_x$ limit for which the
dynamic was originally derived.

\subsubsection{General solution}
Returning to the clamped polymer in the presence of some
background flow, we project the definitional statement
$
h\left(\alpha,\tau\right)=g\left(\alpha,\tau\right)+{\bar h}\left(\alpha\right)
$
onto the $\PHI_k\left(\alpha\right)$:
\beq
h_k\left(\tau\right)=g_k\left(\tau\right)+{\bar h}_k,
\eeq
which implies the initial condition
$g_k\left(0\right)=h_k\left(0\right)-{\bar h}_k$.
Recalling the simple time-dependence of the modes $g_k$
from Eq.
(\ref{gsol}), we see
\beq
h_k\left(\tau\right)={\bar h}_k\left(1-{\rm e}^{-k^4\tau}\right)
+h_k\left(0\right){\rm e}^{-k^4\tau}.
\eeq
The dynamic thus mimics that of a capacitor, charging up with the
final shape-state and draining of the initial shape-state,
each mode governed independently
with decay rate
$k^{-4}$. In real space,
\ba
h\left(\alpha,\tau\right)&=&{\bar h}\left(\alpha\right)
+\sum_{k=k_1}^\infty\PHI_k\left(\alpha\right){\rm e}^{-k^4\tau}\nonumber \\
&& \times
\int_0^1\! d\alpha'\PHI_k\left(\alpha'\right)
\left[h\left(\alpha',0\right)-{\bar h}\left(\alpha'\right)\right].
\label{EHDIrs}
\ea
In the experiment considered, the initial
condition is a flat polymer: $h\left(\alpha,\tau=0\right)=0$.
Since ${\bar h}$ is the solution to
${\bar h}_{\alpha\alpha\alpha\alpha}=1$, with boundary conditions
${\bar h}\left(0\right)=
{\bar h}_{\alpha}\left(0\right)=
{\bar h}_{\alpha\alpha}\left(1\right)=
{\bar h}_{\alpha\alpha\alpha}\left(1\right)=0$,
we find upon integrating by parts
that
\beq
\int_0^1d\alpha'\PHI_k\left(\alpha'\right){\bar
h}\left(\alpha'\right)={{\bar\PHI_k}\over k^4}~,
\eeq
where
$ {\bar\PHI_k}\equiv \int_0^1d\alpha\PHI_k$,
and thus Eq. \ref{EHDIrs} reduces to
\beq
h\left(\alpha,\tau\right)
=
{\bar h}\left(\alpha\right)-
\sum_{k=k_1}^\infty
\PHI_k\left(\alpha\right)
{{\bar\PHI_k}\over k^4}
{\rm e}^{-k^4\tau}
{}.
\eeq

Evaluating the first few integrals, we find for
${\bar h}_k\equiv \int d\alpha\PHI_k{\bar h}={\bar \PHI_k}/k^4$,
\ba
{\bar h}_{k_1}&\simeq&6\times10^{-2}~, \ \ \
{\bar h}_{k_2}\simeq 9\times10^{-4}~, \ \ \
{\bar h}_{k_3}\simeq 7\times10^{-5}~, \nonumber \\
{\bar h}_{k_4}&\simeq& 1\times10^{-5}~ \ \ \ \ldots
\ea
Each mode with $k > k_1$ decays exponentially faster
in time than the lowest mode, which thus
dominates as $\tau\goto\infty$, so
\beq
h\goto {\bar h}(\alpha)-{\bar h}_{k_1}{\rm e}^{-k_1^4\tau}
\simeq{\bar h}-0.06\PHI_{k_1}{\rm e}^{-12.36
\tau}.
\eeq

Our picture of the impulsive dynamic of elastica in viscous flow
is thus as follows: we project onto a special function space
in which the long-time solution and the difference between
initial data and the long-time solution
exponentially charge and decay, respectively,
each mode behaving independently.
We are left with only the the long-time solution as the asymptotic limit
$\tau\goto\infty$.

\section{Elastohydrodynamic problem II}

In Stokes II, the driving force is exerted by a wall oscillating
with velocity ${\bf u}=U\ey\cos\left(\omega t\right)$, or position
$y=y_0\cos\left(\omega t\right)$.
To solve
the steady-state limit of SII, we postulate $u\left(x,t\right)=U\Re
\left(e^{i\omega t} G\left(\eta\right)\right)$, where
$\eta=x{(\omega/\nu)}^{1/2}$.
Inserting into \rref{diff_eqn}, we then see $G$ satisfies
\begin{equation}
i G=G_{\eta\eta}~,
\label{ode_stokesII}
\end{equation}
for which the solution vanishing as $\eta\to \infty$ is
$G={\rm e}^{-\sqrt{i}\eta}$.  We then find $u\left(x,t\right)=U{\rm
e}^{-\eta/\sqrt{2}}
\cos\left(\omega t-\eta/\sqrt{2}\right)$, or, in a form useful for
comparison to the elastohydrodynamic case,
\beq
u\left(x,t\right)=U{\rm e}^{-S_4\eta}
\cos\left(C_4\eta-\omega t\right)\label{SIISoln}
\eeq
where $C_4=\cos(\pi/4)$ and $S_4=\sin(\pi/4)$.
This solution describes right-moving waves of velocity
$\omega\ell/C_4$, decaying as $x\goto\infty$ with decay length $\ell/S_4$.

We now consider  a polymer held by
an optical trap which moves with
position $y\left(x=0\right)=y_0\cos\left(\omega t\right)$.
Since we have shown in the previous section
that all modes satisfying the homogeneous
equation of motion with homogeneous boundary conditions decay
exponentially, we must only find a solution in the presence
of inhomogeneity (here, the driving) to find the long-time
limit of the dynamic.
\label{teaser}

In order to verify the validity of our analysis as
well as the plausibility of EHDII as a method for
measuring biopolymer rigidity, we conducted the experiment
and analyzed image data as described below. A scaling relation
predicted by the analysis was confirmed, and a new method for
measurement
of the persistence length of actin was demonstrated.

\begin{figure} \epsfxsize=3.0 truein
\centerline{\epsffile{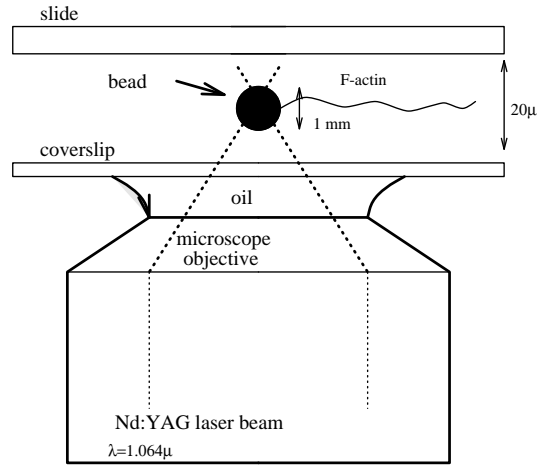}} \smallskip 
\caption[]{Experimental setup.} 
\label{setup} \end{figure}

The experimental setup is shown in Fig. \ref{setup}: F-actin is
bound to a silica bead which is optically trapped.
By oscillating the position of the bead sinusoidally in time,
the filament wiggles back and forth, propagating waves
of displacement down its length. The motion relative to the fluid
is opposed by the fluid viscosity,  and the ``wiggles" are opposed
by the elasticity of the polymer.

\dofloatfig  \fi
\subsection{The characteristic length}
\label{guess}

The elastic constant $A$ has units of
${\rm energy}\times {\rm length}$,
while the viscous force per unit length per unit velocity has the
dimensions of a viscosity or action density $\mu$:
\beq
[\zeta]=[\mu]={{\rm mass}\over {\rm length}\times {\rm time}}=
{{\rm energy}\times {\rm time}\over {\rm length}^3}~.
\eeq
Thus, the natural length obtained from $A$, $\zeta$, and the frequency of
oscillation $\omega$ is
\beq
\ell\left(\omega\right)=\left({A\over \omega \zeta}\right)^{1/4}=
\left({k_B T L_p \over \omega \zeta}\right)^{1/4}.
\eeq
{\it Nota bene} that $\lo$ is {\it not} a mere rescaling of
the persistence length.

With a previously-published persistence length for actin of
$L_p\simeq 15 \mu$m\cite{Ott}, a viscosity $\mu=0.01$ cp,
$k_BT\simeq
4\times 10^{-14}$ erg at $T=300$ K, and measuring
$\omega$ in units of
sec$^{-1}$,
we obtain
\beq
\ell\left(\omega\right) \simeq \left(2.8{\mu {\rm m}\over s^{1/4}}\right)
\omega^{-1/4}.
\eeq
Thus for frequencies on the order of $1$ Hz, we obtain length scales of order
microns, somewhat {\it below} the persistence length.
This range of frequencies seems quite advantageous for experiment.

This elastohydrodynamic length
$\ell\left(\omega\right)$ is precisely the length
found upon nondimensionalizing
the equation of motion (\ref{eqmotdim}).
By analogy to SII, we define the dimensionless
coordinate $\eta=x/(\nut \omega)^{1/4}=x/\lo=x\left(A / \omega\zeta
\right)^{-1/4}$ and
rewrite the solution as
\beq
y\left(x,t\right)=y_0{\Re}\{{\rm e}^{i\omega t} h\left(\eta\right)\}~
\label{defh}
\eeq
and \rref{hyperdif} as
\beq
ih(\eta)=-{\partial}_\eta ^4 h\left(\eta\right).
\label{emn}
\eeq
The solutions of (\ref{emn}) are of the form
\beq
h\left(\eta\right)=c {\rm e}^{\gamma \eta}
\eeq
where $\gamma$ may be any one of the four distinct (complex) numbers such
that $ \gamma^4=-i$.
These are
\beq
\gamma_j=i^j{\rm e}^{-i \pi /8} \ \ \ \ \ \ \ \left(j=1\ldots 4\right)~.
\eeq
The general solution is the sum of these four solutions,
\beq
h\left(\eta\right)=\sum_{j=1}^4 c_j{\rm e}^{i^j z_0 \eta}~,
\eeq
where
\beq
z_0\equiv {\rm e}^{-i \pi /8}\simeq 0.92 - 0.38~i~.\label{z0def}
\eeq
The unpleasant (but certainly not subtle) remainder of the problem
is to solve for the four $c_j$'s,
given some four boundary conditions. At the left end, we enforce not
only the position at $x=0$ but the condition of torquelessness, as
appropriate for an optical trap, $y_{xx}(0)=0$. The right end
must satisfy the free end boundary conditions, \rref{bclin}.
The $c_j$ derived from these conditions are
functions of a rescaled polymer length
${\cal L}\equiv  L/\ell\left(\omega\right)$ and may properly
be written as $c_j\left({\cal L}\right)$.

\subsection{Semiinfinite polymer}
\label{inf_sec}
The
exact solution for $h\left(\eta\right)$ is presented in
Appendix \ref{exactsolnapp}; it simplifies greatly, however,
for
extreme values of $L/\lo$.
In the limit of an infinitely long polymer, the two coefficients
$c_j$ for which $\gamma_j$ has a nonnegative real part must
be zero, allowing only decaying solutions as $x\goto\infty$.
\dofloatfig \begin{figure} \epsfxsize=3.3 truein
\centerline{\epsffile{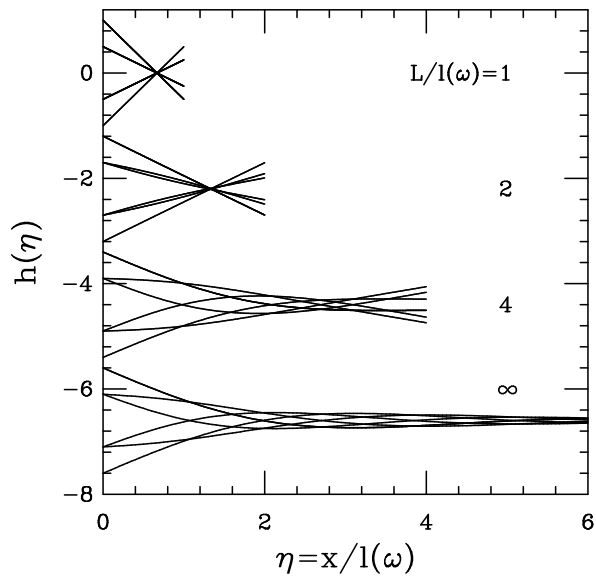}} \smallskip 
\caption[]{Solutions to EHD problem II for filaments of various 
rescaled lengths ${\cal L}$.} \label{fig_finite} \end{figure} \fi

The solution consistent with the two left-end boundary
conditions is
\ba
y=&{y_0\over 2}\Bigl[&{\rm e}^{-C_8 \eta} \cos\left(S_8 \eta + \omega t\right)
\nonumber \\
&+&{\rm e}^{-S_8 \eta} \cos\left(C_8 \eta - \omega t\right)\Bigr],
\label{semi_soln}
\ea
where $C_8=\cos\left(\pi/8\right)$ and $S_8=\sin\left(\pi/8\right)$.
Compare
with the solution to SII, \rref{SIISoln}.
The semiinfinite solution (\ref{semi_soln}) is shown at the bottom
of Fig. \ref{fig_finite} for
$\omega t = n  2 \pi / 6$, $n=1\ldots 6$.
In the hydrodynamic case, the solution \ref{SIISoln} describes
exponentially decaying right-moving traveling waves of transverse
velocity. In the elastohydrodynamic case, the higher-order
derivative allows more complicated behavior: right
and left-moving waves of displacement, with different decay
rates and velocities. In this case, the right-movers have a slower decay
(since $S_8\simeq 0.38< 0.92\simeq C_8$),
and might be expected
in some sense
to dominate  over the left-movers.
This mechanism will be elaborated on in section \ref{force}.

\subsection{Finite polymer}
\label{short_sec}
In the limit of a short or stiff polymer, ${\cal L} \ll 1$, we may
expand the coefficients $c_j\left({\cal L}\right)$ appearing
in the exact solution to find
\ba
h_{\ll}\left(\eta\right)  &\simeq  \left({z_0 {\cal L}}/8\right) & \left[
\sin\left(z_0 \eta\right)-  \sinh\left(z_0 \eta\right)\right]
\nonumber \\
 & + \left(1/2\right)  & \left[  \cos\left(z_0 \eta\right)+  \cosh\left(z_0
\eta\right)\right] \nonumber \\
& - \left(3 /4 z_0 {\cal L}\right) & \left[  \sin\left(z_0 \eta\right)+
\sinh\left(z_0 \eta\right)\right]~.
\label{smalltrig}
\ea

The apparently singular result $h \sim {\cal L}^{-1}$ is
shown illusory by rewriting $\eta=\alpha {\cal L}$, $\alpha
=x/L \in \left(0,1\right)$ and expanding
(\ref{smalltrig}) for small ${\cal L}$, yielding
\ba
h_{\ll}\left(\alpha\right)&\simeq& \left(1-{3 \over 2}\alpha\right)\nonumber \\
&&+{i \alpha {\cal L}^4 \over 1680}
\left(-16+70\alpha^2-70\alpha^3+21\alpha^4\right)~.\label{short_soln}
\ea
Equivalently, we may derive this polynomial by
truncating a series expansion for $h$ in $\alpha$ and
enforcing the equation of motion (\ref{emn}) and the
boundary conditions (\ref{bclin}).

Using $h_{\ll}\left(\eta\right)$ (in \ref{smalltrig}),
the boundary conditions at $x=0$ and
\rref{emn} are
satisfied exactly, whereas the boundary conditions at $x=L$
are satisfied to order ${\cal O}\left({\cal L}^{5}\right)$.
Using $h_{\ll}\left(\alpha\right)$ (in \ref{short_soln}),
all four boundary conditions are satisfied exactly, whereas \rref{emn}
is solved to order ${\cal O}\left({\cal L}^{4}\right)$.

The exact solution is shown in
Fig. \ref{fig_finite} for ${\cal L}=1, 2, 4$, and $\infty$ and
$\omega t = n  2 \pi / 6$, $n=1\ldots 6$.
Note the existence of a pivot point at $x={2L/3}$ as
${\cal L}\goto0$. This behavior is described by the
${\cal O}({\cal L}^0)$ term in \rref{short_soln}: as
${\cal L}\goto 0$, the polymer acts as a rigid rod.
As a consequence, it is impossible to tell if a movie
of such a polymer is being played forward or backward.
Indeed, this is a filamentous version of the famous ``one-armed swimmer"
or ``scallop" example illustrating the lack of
net propulsion for rigid objects executing time-reversible motions in
low Reynolds number flow \cite{purcell,childress}.

\subsection{Propulsive force}
\label{force}
Problem II and its associated experiment
are sufficiently reminiscent of
flagellar hydrodynamics to motivate a calculation of the
propulsive force $F$ generated in the $x$ direction by the wiggling.
This can be done by integrating ${\bf f}_{\cal E}$, the
force exerted by the polymer on the fluid,
along the length of the filament. We then contract
this instantaneous total
force with $\ex$ and average over one period.
This force is equal and opposite the propulsive force exerted by the
fluid on the polymer.

Noting that the force per unit length in \rref{exact_bcs}
is a total derivative,
\beq
{\bf f}_{\cal E}=A\partial_s\left(\kappa_s{\n}-\oh\kappa^2\ta\right),
\eeq
and recalling the boundary conditions imposed on $\kappa$ and $\kappa_s$, we
have
\ba
F\equiv-\int ds~{\bf f}_{\cal E}\cdot \ex&=&
-A\ex\cdot\left(\kappa_s{\n}-\oh\kappa^2\ta\right)\biggl|_0^L\\
&=&A\ex\cdot\left(\kappa_s{\n}\right)\left(s=0\right)\\\nonumber
&=&A\kappa_s\sin\theta\left(s=0\right).\nonumber
\ea
This is geometrically exact.
We now wish to calculate the time average over one period,
which we denote by $\langle\cdots\rangle$.
Within the linearized solution,
$\kappa_s\sin\theta\simeq y_{xxx}y_x$.
Recalling the expression for $y$ in \rref{defh}, and expressing
the answer in terms of the rescaled variable $\eta$, we obtain
\ba
F&=&A\langle y_x y_{xxx}\rangle\\
&=&{Ay_0^2\over2\lo^4}\Re\left(h_{\eta}h_{\eta\eta\eta}^*\right)\\
&=&{y_0^2\zeta\omega\over 4\sqrt{2}}\Upsilon\left({L\over \ell(\omega)}\right)
\ea
where $^*$ indicates complex conjugation, $\lo$ is the characteristic length,
and $\Upsilon$ is a scaling function conveniently normalized (see below).

\begin{figure} \epsfxsize=3.0 truein
\centerline{\epsffile{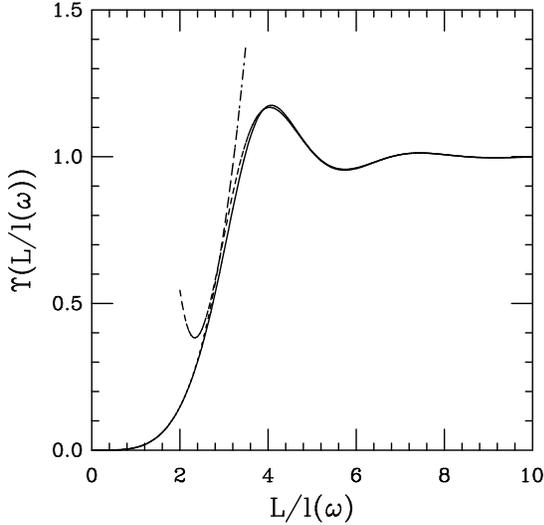}} \smallskip 
\caption[]{Scaling function $\Upsilon$ for propulsive force. 
The large ${\cal L}$ expansion is plotted for ${\cal L}>2$, 
and the small-${\cal L}$ solution is plotted for ${\cal L}<3.5$.} 
\label{force_fig} \end{figure}

The exact solution to EHDII given in Appendix \ref{exactsolnapp} can be used to
calculate
the function $\Upsilon$ for all values of the polymer length.
The asymptotic behavior  as ${\cal L}\goto\infty$ is
\beq
\Upsilon({\cal L})
\goto 1+
4{\rm e}^{-2S_8{\cal L}}\sin\left(2C_8{\cal L}\right).\label{longforce}
\eeq
When the length is short compared to the characteristic
length, so the polymer flexes very little,
\beq
\Upsilon \simeq
{11\over3360}{\cal L}^4
+{\cal O}\left({\cal L}^8\right)
\simeq {11\over3360}{\zeta\omega\over A}L^4.\label{shortforce}
\eeq
As Fig.
\ref{force_fig} illustrates,
the short-length approximation (\ref{shortforce})
shows good agreement with the
exact solution for
${\cal L}\lesssim 3$, as does the large-${\cal L}$ approximation
(\ref{longforce})
for ${\cal L}\gtrsim 3$. The approach to the asymptotic limit is
oscillatory, with a maximum near ${\cal L}\sim 4$,
the value at which $\Re\{h_{\eta}\}$ acquires
its first root, and a local minimum near ${\cal L}\sim 6$,
the the value at which $\Im\{h_{\eta}\}$ acquires its second root.
The unexpected local maximum indicates that there is an
optimal combination of $A,\omega$, and a finite $L$.

\dofloatfig  \fi

Inserting typical numbers from the experiment (in cgs),
\begin{eqnarray}
\mu     \sim & 10^{-2}, ~ y_0~&  \sim  4 \times 10^{-4}\nonumber\\
\omega  \sim & 2\pi,	~ {L\over d}&
\sim {7\times 10^{-4}\over7 \times 10^{-7}}=10^{3}
\end{eqnarray}
we find ${F(\infty)}\sim 2\times 10^{-9}$ dynes $=3\times 10^{-2}$ pN.
For a trap stiffness of $\sim .02$ pN/nm, this would induce a displacement
of $1.5$ nm, at the lower limit of experimental observation.
The production and measurement
of propulsive
force by an artificial flagellum
was attempted by G. I. Taylor \cite{git}
using a glycerine-filled tub to mimic the
low Reynolds numbers found {in~vivo}. Taylor struggled
to drive the flagellum without inducing unwanted torque or
disturbing the flow, a difficulty obviated
by the use of optical traps.

Returning to the asymptotic expressions for $h$ derived
in the sections \ref{inf_sec} and \ref{short_sec},
we observe a pleasant accordance with
the qualitative features of Fig. \ref{force_fig}. In the semiinfinite
case, we noted the presence of right- and left- movers,
with right-moving waves of displacement exhibiting
slower decay. Such
a dominance accounts for the nonzero propulsive force in the
${\cal L}\goto\infty$ limit, where a net propulsion to the
left survives. In the ${\cal L}\goto 0$ case, we recovered
a shape which asymptotes to a pivoting rigid rod,
not unlike a one-armed swimmer. As we expect from life
at low Reynolds number \cite{purcell,childress}, such a motion,
invariant under $t\goto -t$, can produce no net propulsion.

As further illustration of the relationship between
low-Reynolds-number swimming and cyclic motions, we
observe that the lowest-order expression for the time-averaged
force is equal to
\beq
F={\zeta\omega\over 2\pi}
\int_0^{2\pi/\omega} dt~y_x\vert_{x=0}\partial_t\int_0^L dx~y(x)
\eeq
or, noting that $\int dx~ y(x,t)$ is simply the area $\A(t)$ under the
curve $y(x,t)$, and that the slope at the left is to first order
simply the tangent angle $\theta_0$
\ba
F&=&{\zeta\omega\over 2\pi}\int_0^{2\pi/\omega} dt ~\theta_0~ {d\A \over dt}\\
 &=&{\zeta\omega\over 2\pi}\oint \theta_0 d\A.
\ea
This result can be interpreted
quite simply: the propulsive force results from pushing aside some volume
(or in two dimensions an area) of fluid, projected in the
direction of propulsion $\ex$ an amount proportional to $\theta_0$.
Note that had we been interested in the propulsion in the
transverse ($\ey$) direction, the $\theta_0$ would not appear, leaving
the absence of net forcing: $F\propto\oint d\A=0$, as we would expect.

The net force, then, is the area enclosed by a trajectory in
$\A - \theta_0$ space during some cyclic motion. This
representation is independent of the particular motion exhibited, although
we have here considered simple periodic motion, for which
the trajectory is always an ellipse. As ${\cal L}\goto 0$,
the elliptical trajectory thins to
a straight line, encloses no area, and thus produces
no force, as commented on above.

This representation makes clear that
in an inertialess world, net motion
is principally geometric in origin rather than dynamic.\cite{frank}
In a manner analogous to the importance of path rather
than kinetics in generating net work in a Carnot diagram, we see
that we can remove time entirely from the expression and
consider instead a path in a
low-dimensional projection of the infinite-dimensional
shape space.

\section{Experiments on actin}
As mentioned in section \ref{teaser}, the EHDII experiment was performed
and the data compared to the solution of \rref{emn}.
In this way we were able to confirm the results of the
analysis and investigate a new
method for measuring biopolymer
bending moduli.

\subsection{Materials and methods}

Actin from white leghorn chicken breast was purified after a published
procedure\cite{Pardee}. Filamentous actin (F-actin) was fluorescently labeled
with rhodamine-phalloidin (R-415, Molecular Probes, Inc.).
Anti-actin antibodies (Sigma) were covalently coupled through carbodiimide
to fluorescent polystyrene
beads (CX, Duke Scientific Corporation),
of diameter 1 mm,
after the recommended protocol of Duke. Experiments were performed in
actin suspension buffer containing 25 mM imidazole, 25 mM KCl,
5 mM b-mercaptoethanol at pH 7.65. To avoid photobleaching during observation,
1mg/ml glucose, 33 units/ml glucose oxidase, 50 units/ml catalase were added
to the suspension buffer.

In order to prevent actin filaments from sticking to the glass surfaces,
slides were coated with bovine serum albumin (BSA). Observation chambers
were sealed with nail polish and filled by capillary action. Beads and F-actin
coupled within the cell; the ratio beads:F-actin was adjusted to have one
or two beads per filament. Beads bound to various locations along filaments.
Among these filaments, we chose bead-ended actin filaments for the experiment.

\begin{figure} \epsfxsize=3.6 truein
\centerline{\epsffile{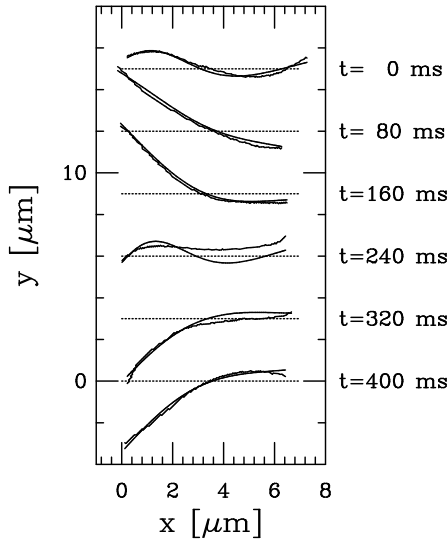}} \smallskip 
\caption[]{Series of typical images of driven actin filaments in 
EHDII. Also shown are fits to the solution of \rref{emn}.} 
\label{Typ_Image} \end{figure}

Observations were made on a Zeiss microscope (Axiovert 135) equipped
with a 100 W mercury lamp and a standard filter set (XF37, Omega optical).
To prolong observation time, the excitation light intensity was reduced by
inserting neutral density filters. Fluorescent images were taken through a
4X TV tube, via an image intensifier (Hamamatsu) followed by a CCD camera
(XC-77,
Sony). Images were recorded with an S-VHS recorder. Images corresponded to an
overall screen size of 44 mm in horizontal dimension leading to a value of
0.06mm/pixel after digitization. Fig. \ref{setup}
shows the experimental setup,
and Fig. \ref{Typ_Image} shows a series of typical images. The smooth
curves indicate fit solutions to \rref{emn}.

\dofloatfig  \fi

The optical tweezer was made by focusing a 0.5 W Nd:YAG (Spectra) laser
beam through a Zeiss 63X, 1.4 numerical aperture Plan-Apochromat microscope
objective.
A mirror mounted on a galvanometer controlled by a function generator
served to
move
trapped
beads in the focal plane (Fig. \ref{setup}). We imposed
sinusoidal driving, with frequencies and amplitudes ranging, respectively, from
0.1 Hz to
6 Hz and from 5 $\mu$m to 10 $\mu$m.

Up to 10 images per oscillation period were digitized with a Scion frame
grabber
card and a Power Macintosh; they were subsequently analyzed using NIH-Image
software.
Each image was retraced by mouse
and its coordinates  determined.
Only those sequences
for which
filaments remained in the focal plane for
an oscillation period
were retained for data analysis.
\label{exptdet}

\subsection{Experimental results}

\begin{figure} \epsfxsize=3.3 truein
\centerline{\epsffile{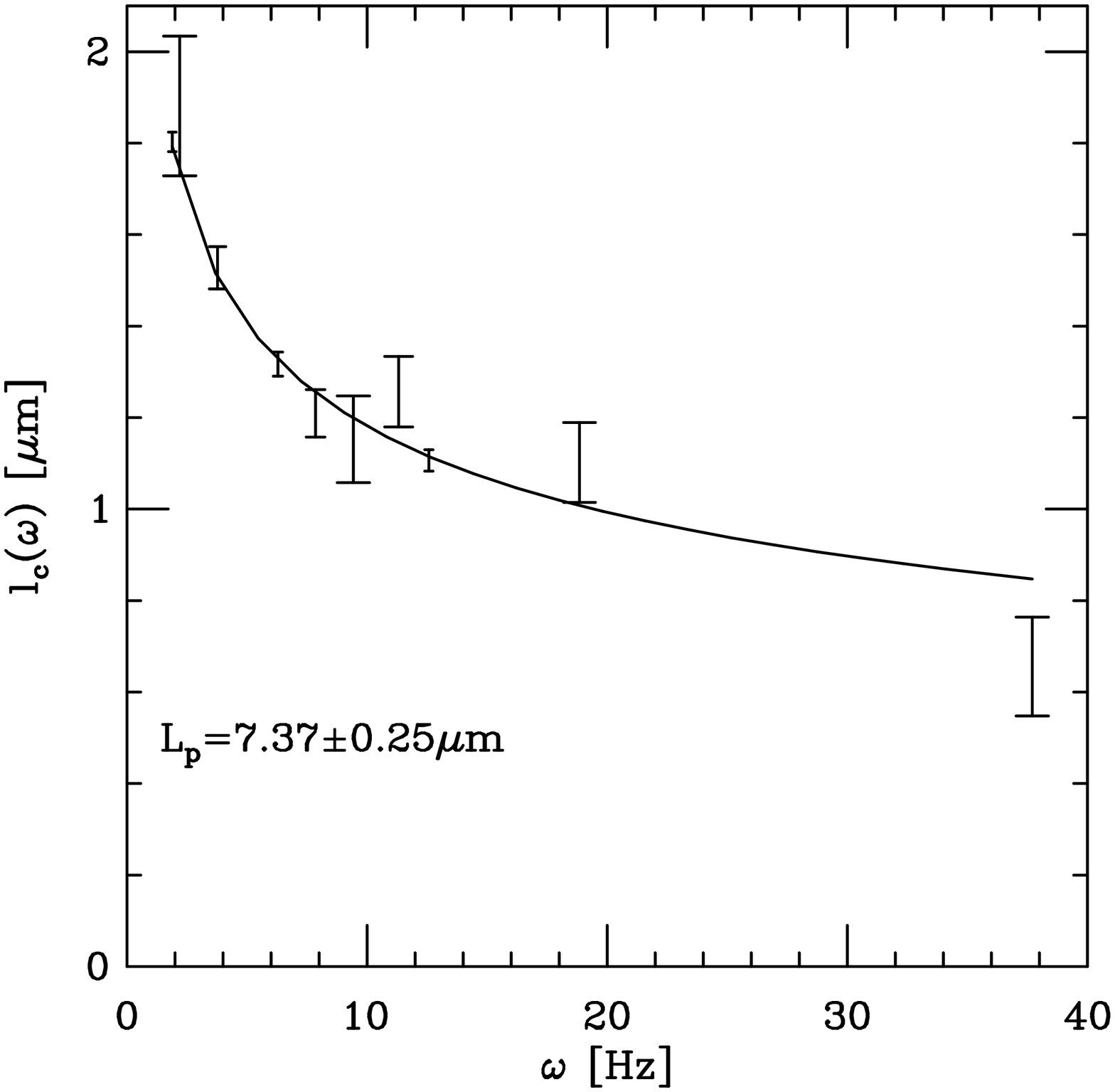}} \smallskip 
\caption[]{The characteristic length scale $\lo$ vs. frequency $\omega$.  
The smooth curve is a fit to \rref{fit} for $L_p=7.37 \mu$m.  
The scaling of the characteristic length with the fourth root of time 
suggests that actin is well-described by Eq. \ref{hyperdif} and thus 
is a semiflexible polymer, with a scale-independent elasticity.} 
\label{l_of_omega_fig} \end{figure}

Knowing the amplitude of the driving of the bead ($y_0$) and
the frequency ($\omega$), and reading off the projected length ($L$) and
the phase ($\omega t$) directly from the images,
we were left with a {\it one-parameter}
fit of the images
to the solution of \rref{emn},
varying only $\lo$ to minimize $\chi ^2$.
We can then observe the dependence of $\ell$ on $\omega$, as  illustrated
in Fig.
\ref{l_of_omega_fig}. The variation in error bars
can be attributed to the widely-varying numer of images taken
at different frequencies.

Comparing with the earlier analysis (\cf~ section \ref{guess}),
we can extract from this scaling a measurement of the persistence
length. Fitting to
\beq
\lo={\left(k_BT\over\zeta\right)}^{1/4} L_p^{1/4} \omega^{-1/4},\label{fit}
\eeq
we determine $L_p$ to be $7.37\pm0.25\mu m$.

\dofloatfig  \fi

\subsection{Discussion}
\label{discussion-sect}

There are a few limitations with
this realization of the experiment
which, upon correction, will improve this technique and make
the data more conclusive.
An obvious mechanism for improving the error bars is to
accumulate more data. Image-taking was entirely manual,
with hundreds of student-hours spent looking for usable
images. An automation of this process would clearly
be advantageous and improve the low statistics used
here.

With careful control of the timing,
images of equal phase can be superposed to
average out thermal fluctuations or experimental variation
in the images before fits are performed.

Most importantly, since our aim was to verify the plausibility of the
experiment,
we did not limit ourselves to small-amplitude wiggling, thus
leaving the realm of validity of the small-$y_x$ approximation.
We can estimate the error due to such driving by looking
at the relevant terms from the geometrically exact equation of
motion:
\beq
{y_t\over{({1+y_x^2})^{1/2}}}
=-\nut\left({y_{xx}\over{\left(1+y_x^2\right)^{3/2}}}\right)_{ss},
\eeq
If we wish to approximate this with Eq. \ref{eqmot},
we are measuring an ``effective" $\nut$, or $\enut$,
where
\beq
\enut\simeq{\nut\over\langle \left(1+y_x^2\right)^2\rangle },
\eeq
where the brackets indicate averaging over the data,
and thus the true $A$ will be underestimated by a factor
of $\langle({{1+y_x^2}})^2\rangle$, which is always greater than unity.
Inspecting Fig. \ref{Typ_Image}, we
see that there are data for which $y_x$ is not necessarily small.
For this reason, the data we collected can only put a lower
bound on $A$.
We anticipate that the true value may be greater by a factor up
to $\sim 1.5$; future experiments clearly should employ smaller-amplitude
driving.

\subsection{Suggestions for further analysis}

We also anticipate that a more accurate treatment
of the geometry and hydrodynamics would
refine the technique. The true geometry is nonlinear and hydrodynamics
nonlocal, but neither intractable and both amenable to
numerics.
The geometrically exact, intrinsic formulation involves some
enjoyable mathematics of curve dynamics, \cite{herbie}
whereas the linearized treatment presented herein is more
illustrative and more easily connected with experiment.
Similarly, in an attempt to make the analysis as clear
as possible, we have omitted from \rref{eqmot} the
background flow due to the trapped bead.

Some amount of discussion has been entertained in the biophysical
community about the possible scale- or time-dependence of the
elasticity of biopolymers, including both actin \cite{Kas}
and microtubules \cite{Kurachi}. One of the more powerful
features of this technique is that, since it involves a
controlled dynamic,
specific scales and frequencies can
be investigated to
attempt a spectroscopy of elasticity.
Equation \ref{eqmot} can be extended without difficulty to
include a characteristic relaxation time $\tau$ or a
continuum of
times,
in an attempt to model a characteristic rate of bond-breaking
in the presence of bending. Similarly, one can include
additional bending moduli which depend on higher-order derivatives.
For oscillatory motion, occurrences  of $\partial_t$ are simply
replaced by $\partial_t+{1/ \tau}$ or ${i\omega+1/\tau}$.
Including higher-order derivative
terms simply results in
replacing $A y_{2x}$ with
$Ay_{2x}+By_{4x}$ in the bending energy
and $Ay_{4x}$ with $Ay_{4x}+By_{8x}$ in the
equations of motion.
We then may recover such an equation as
\beq
\zeta(\p_t+{1\over \tau})y= -Ay_{xxxx}-By_{8x},
\eeq
which may be solved in the manner of
Eq. \ref{hyperdif}. This more general
expression makes possible the experimental confirmation or
refutation of such hypothesized mechanisms.

\section{Comments on previous experiments}

A recent pair of elastohydrodynamic experiments
involving microtubules \cite{felg},
brought to our attention as the original
version of this paper was being completed,
provides an excellent
opportunity to apply the spirit of analysis which
we have developed for EHDI. In both, the crucial experimental
observable is the motion of the free end of a microtubule;
the analysis must then relate this motion to the bending
modulus $A$.
We will first investigate
the analysis appropriate to a special case of EHDI in which
a polymer relaxes to a straight configuration in the absence
of a driving flow. We then investigate a more complicated
experiment in which an optical trap induces a force
in the middle of the polymer.

\subsection{The Simple EHDI experiment}

In the first experiment \cite{felg}, a microtubule is initially clamped at
the left to an axoneme and
trapped directly at the free end;
the polymer  is bent out of its mechanical equilibrium configuration. When
the trap is shut off, the polymer relaxes back to the
straight shape in a way which
we may describe as before: the initial condition is
projected onto the appropriate space, in which each mode
decays independently.

\dofloatfig \begin{figure} \epsfxsize=2.8 truein
\centerline{\epsffile{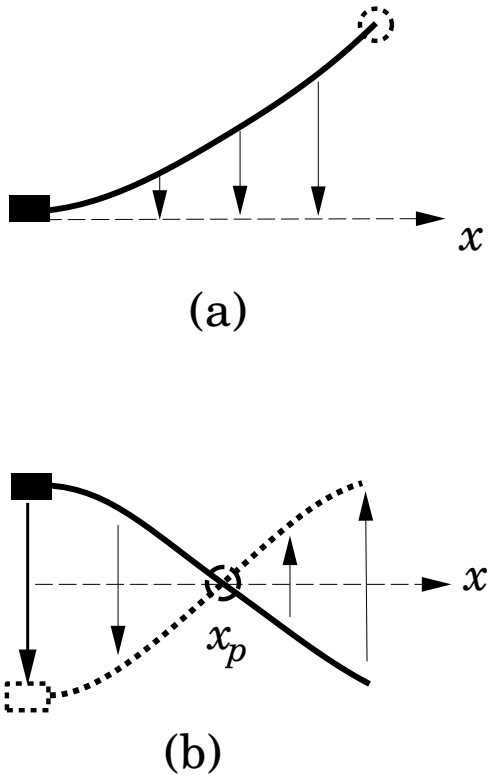}} \smallskip 
\caption[]{EHDI experiments. (a) a simple special case (b) 
end-driving as well as driving via a point force.} \label{seesaw_fig} 
\end{figure} \fi

What remains then is merely to determine the initial data: the shape
of a biopolymer clamped at one end and held by a trap at the other.
We may derive this from
the (geometrically exact) general equations of force
and torque balance for the {elastica}:
\ba
&{\bf N}_s+{\bf f}_e=0\label{fbal}\\
&{\bf m}_s+\left({\bf r}\times {\bf N}\right)_s+{\bf r}\times {\bf
f}_e=0\label{tbal}
\ea
Here ${\bf N}$ is a contact force (not the
unit normal),
${\bf m}$ is a bending moment, and ${\bf f}_e$ is some external force.
While these equations may appear unfamiliar, they in fact have
a long history as the
classical forms of
the equations of equilibrium for the special theory of Cosserat
rods \cite{antman}.
Combining the equations into one,
\beq
{\bf m}_s=\ta\times\int_0^sd\sigma~{\bf f}_e\left(\sigma\right)~,\label{shape}
\eeq
where $\ta={\bf r}_s$ is the unit tangent.
In two dimensions, cross products are scalars, and
${\bf m}_s$ becomes
the scalar $m_s$.

We now include the defining constitutive relation of linear
elasticity:\footnote{Here, ``linear" refers to the curvature in the
constitutive relation, not necessarily in the equations of motion,
in which the geometry is often the source of nonlinearity.} $m=A\theta_s$.
Furthermore, in considering {\it point} external forces
(the axoneme and the trap) that
act at the left and right ends, we write
\beq
{\bf f}_e(s)={\bf F}_0\delta(s)+{\bf F}_1\delta(s-L)~.
\label{pointforces}
\eeq
We assume that there is no net
force, so ${\bf F}_0=-{\bf F}_1$.
Noting that $\ta=\left(\cos\theta,\sin\theta\right)$ we rewrite
\rref{shape} (for $0<s<L$) as
\beq
A\theta_{ss}=-F_y\cos\theta +F_x\sin\theta \label{nwpe}
\eeq
where $F_{\{x,y\}}$ are the components of the
force exerted
{\it by the trap}, rather than by the axoneme.
We expect $F_x<0$ and $F_y>0$.
\rref{pointforces} is geometrically exact and can
be solved in terms of elliptic integrals.

Note that there is no
reason to assume the trap exerts a force only in the $\ey$ direction --
the oft-quoted ``cantilevered beam" problem from introductory
civil engineering
texts. We show below, however, that the $\ex$ component
is higher order in $\theta$ and will be ignored in the linearized
treatment presented.
Linearizing Eq. \ref{nwpe} for small $\theta$ obtains
\beq
A\theta_{ss}=-F_y+\theta F_x,\label{wpe}
\eeq
the windy pendulum equation\cite{me}.
Assuming the polymer is homogeneous along its arclength, we note
that there is no energy cost upon moving the trap along
the axis of the polymer. Therefore the trap can exert no force
in the tangential direction. We may write this constraint
\beq
0={\bf F}\cdot {\bf {\ta}}
\simeq F_x+F_y{dy\over dx}~.\label{TFisN}
\eeq
Accordingly, $F_x/F_y\simeq - dy/dx$,
and \rref{wpe} becomes
\ba
Ay_{xxx}\simeq  -F_y\left(1 + {\cal O}(y_x^2)\right)~.
\label{wpel}
\ea

We now must consider the boundary conditions.
The axoneme clamps the left end of the polymer, thus
$\nobreak{y\left(0\right)=y_x\left(0\right)=0}$.
Since there is no energy cost to rotating a polymer
held in an optical trap, there should be no bending moment, implying
$y_{xx}\left(L\right)=0$.
Given these three boundary conditions, the solution to (\ref{wpel}) is
\beq
y\left(x\right)={{\tilde y}\over 2}\left(3\alpha^2-\alpha^3\right)~,
\ \ \ \ {\tilde y}=y(L)={F_yL^3\over 3 A}~,
\label{InitData}
\eeq
where again $\alpha\equiv  x/L$.

\label{initdata}

Now that we have our
initial data, we project it onto the appropriate function space
in which the dynamics are simply exponential relaxation.
In keeping with the experiment we seek here to model,
we focus on the motion of the free end, whose rate of
relaxation provides a direct means of measuring
the bending modulus $A$. Given the dramatic
increase in relaxation rates for each subsequent mode (\cf~
Eq. \ref{gsol}), we expect only the lowest mode to
be relevant beyond negligible initial times.
Moreover, inspecting Fig. \ref{fte}, we
observe that the eigenfunction $\PHI_{k_1}$ well-approximates the
normalized third-order
polynomial which describes the initial data,
 whereas higher modes more closely resemble Fourier
modes. This close agreement, rather surprising from a sum of
trigonometric and hyperbolic trigonometric functions,
leads to the dominance of the projection of initial data
onto the first mode; specifically
\beq y_{k_1}\left(0\right)\simeq0.971~{\tilde y};~~~
y_{k_2}\left(0\right)\simeq-0.0247~{\tilde y}~,\eeq
where we define $y_{k_n}\equiv\int_0^1d\alpha~y\PHI_{k_n}$,
in analogy to \rref{gsol}.
Inserting typical numbers ($A\sim 5\times10^{-15}$ dynes cm$^2$,
$L\sim10\times10 \mu$m,
and the drag coefficient from
Eq. \ref{dragcoeff} with
$d\sim20 $nm and $\mu\sim.01$ erg s/cm$^3$),
we see that for times beyond $0.01$s,
the amplitudes of subsequent modes are at most 1\% that of the first mode.
The shape is then described by
$
y\left(x\right)\simeq\PHI_{k_1}\left(x\right)y_{k_1}\left(t\right)
$
and decays as ${\rm e}^{-r t}$, with  $r\equiv { A k_1^4 / \zeta L^4}$.
In the model accompanying the experiment, it was assumed
that the shape was described by a single decaying mode for all times
up to $t=0$. As we have shown, $\PHI_{k_1}$ fortunately well-approximates
the initial data such that this introduces only an error ${\cal O}(10^{-2})$.

Within these approximations, the free end
decays as
\beq
{y\left(L,t\right)\over y\left(L,0\right)}\simeq
0.971\PHI_{k_1}\left(L\right){\rm e}^{-rt}
\simeq{\rm e}^{-rt}.
\eeq
so that a measurement of the e-folding time yields $r$ and
hence $A$. It is now clear that the identity of $k_1$, the
solution to the transcendental equation (\ref{solv_cond}), is crucial,
as it is physically manifested in the decay rate.

The model in \cite{felg} is based first on the computation of
the deflection $y_l$ of the free end of the clamped elastica experiencing a
flow linearly increasing in $x$ and constant in time, with
a drag coefficient $\tilde\zeta$.
Such a flow would be appropriate for a rigidly rotating rod
with constant angular velocity, rather than for a bent filament.
Second,
to find the decay rate in \cite{felg}, it is noted
that the exponential relaxation of the tip must correspond
to some first-order differential equation. Although this equation
is not stated, we must assume it to be $y_t(t)=-(v/y_l)y(t)$, (where
$v$ is the maximum flow velocity used in the computation of $y_l$) in order
to recover the reported decay rate,
\beq
r={120\over 11} {\Afelg \over \tilde\zeta L^4}.\label{rodrlx}
\eeq
where $\Afelg$ refers to the value of $A$ which would be
extracted from data, using the simplifications described
above.

A careful analysis of the drag coefficient in
slender-body hydrodynamics
employs a matched asymptotics for the fluid velocity
and is dependent on the shape of the slender body.
Treating a microtubule as a cylinder, the appropriate drag coefficient is
\cite{cox1}
\beq
\zeta={4\pi\mu\over\ln\left(L/d\right)+2\ln2-\oh}. \label{cox}
\eeq
Unfortunately, the drag coefficient used in \cite{felg}
is that appropriate to
tangential rather than normal flow, with numerator $2\pi\mu$
(quoting an earlier misquotation
of Doi and Edwards\cite{tony,DnE}) and further
suffers from the replacement of the constant terms in the denominator
of \rref{cox} with $-\ln2$.

We may now compare the results of a differential equation-motivated
analysis with the model of
\cite{felg}.
Given some measured e-folding time $t_*$,
the eigenmode analysis yields the bending modulus $A$ as
\beq
\Aus={\zeta L^4\over t_* }k_1^{-4}
\eeq
whereas the rodlike treatment implies
(from Eq. \ref{rodrlx})
\beq
\Afelg={\tilde \zeta L^4\over t_*} {11\over 120}.
\eeq
Inserting typical numbers from the experiment,
$L\sim 10\mu{\rm m}$, $d\sim .02\mu{\rm m}$,
we see that
\beq
{\Afelg\over \Aus}=1.13{\tilde \zeta\over \zeta}
\simeq 0.709,
\eeq
a systematic underestimate beyond the uncertainties of experiments.

\subsection{The discontinuous EHDI experiment}

A more complicated example of EHDI which was also conducted
includes driving the axoneme,
so that we must incorporate an inhomogeneous boundary condition, and
point forcing by an optical trap \cite{felg}.
In this experiment, the axoneme (attached to the cover slip) is moved with
constant velocity between two extreme positions.  During this motion,
the polymer is constrained by the optical trap to
pass through some intermediate point $(x_p,y_p)$.
{}From the position and
velocity of the free end it is possible to determine the bending modulus.
Enumerating all the relevant
forces, we consider
the force at the left end due to the axoneme, $F_a$,
the point force due to the trap $F_p$, and the force per unit length
due to drag, $-\zeta h_t$ (\cf~ Fig. \ref{seesaw_fig}).
We insert these terms into the equations of
force and torque balance (\ref{fbal},\ref{tbal}) to find the equations of
motion.

\subsubsection{Explicit declaration of forces}

The sum of the external forces per unit length can be expressed
as
\beq
{\bf f}_e:={\bf F}_p\delta\left(s-s_p\right)+{\bf F}_a\delta\left(s\right)+{\bf
f}_d
\eeq
where ${\bf f}_d$ is the drag (Eq. \ref{eqmot}).

Integrating over the delta functions,
and taking the axoneme to be located at $s=0$,
the geometrically exact governing equation (\ref{shape}) becomes
the fiercely complicated integro-differential equation in $\theta(s,t)$
\ba
A\theta_{ss}&=&
\ta\times\biggl[{\bf F}_a+{\bf F}_p\Theta\left(s-s_p\right) \\
&&\qquad+\zeta\int_0^s\!d\sigma\left(\n\n+\beta\ta\ta\right)
\left(\int_0^{\sigma}\!d\sigma'\n\theta_t -{\bf u}\right)
\biggr],\nonumber
\ea
which greatly simplifies for small $y_x$.
Here $\bf u$ is the background flow velocity, in the frame of the
trap, due to the motion of the coverslip. This is a constant
for the problem, which we define to be $-v_c\ey$ so that $v_c>0$.
Note that the axoneme, attached to the coverslip, also moves
with velocity $-v_c\ey$.
We see that point forcing in the middle
of a polymer introduces a qualitative change in
the dynamic: a Heaviside function in the equation of
motion.

\subsubsection{Linearized geometry}
We adopt an expansion in $\theta$, keeping only the first-order
terms. The drag term, as in \rref{EqMotInhom}, simplifies to $\zeta
{\hat{\bf e}}_y(y_t+v_c).
$

Noting, as in \rref{TFisN}, the constraint that the force
on a polymer due to an optical trap must have no tangential
component, the $y$-component of the linearized equation reduces
to
$
A\theta_{ss}\simeq F_a-F_p\Theta\left(s-s_p\right)-\zeta(\int^sy_t+v_c),
$
which upon differentiation and linearization implies
\beq
\zeta (y_t+v_c)=-Ay_{xxxx}-F_p\delta\left(x-x_p\right).\label{eqndisc}
\eeq
This is our working equation, obtained from the linearized second
derivative of the equation of net torquelessness.

Since we do not know the magnitude of $F_p$
{\it a priori}, we must perform a matching of $y^a\left(x\right)$, the
curve described by the anchored end, and $y^f\left(x\right)$, the curve
described by the free end, which solve
\beq
\zeta( y_t^{\{a,f\}}+v_c)=-Ay_{xxxx}^{\{a,f\}}\label{eqaf}
\eeq
subject to matching conditions at the point of forcing.
Horrifying though this may sound, it is an excellent opportunity
to apply the simple ideas developed in earlier sections.
It is also, of course, a chance to treat in a thorough way the
analysis of a biopolymer subject both to drag and to micromanipulation
via some point force and boundary condition.
Moreover, if we wish to claim that the PDE description is the
appropriate analysis for single biopolymer dynamics, we must
be able to apply it  even to such an awkward case.
We will separate the solutions into
the homogeneous and the particular,
and in a procedure that is now familiar, construct the appropriate
function space in which the dynamic is simple.

\subsubsection{Matching and boundary conditions}
Now that we have two fourth-order equations of motion, we
must specify eight matching and boundary conditions.
Inspecting the equation of motion (Eq. \ref{eqndisc}), we see it supports
a discontinuity in $y_{xxx}$; however, $y_{xx}$, described
by an integral over a Heaviside function, is continuous, as
are
$y$ and $y_x$. Moreover, if we wish to describe
an experiment
in which the filament position is constrained at
the point of forcing $x_p$, we have not only the matching condition
$y^{a}(x_p)=y^{f}(x_p)$, but the stronger condition
$y^a(x_p)=y^f(x_p)=y_p$; here we choose $y_p$ to be $0$
without loss of generality. We are thus describing a polymer pinned
at a certain intermediate point along the curve
by an optical trap,
while the right and left sides perform some coupled motion.

At the free end, we impose forcelessness and torquelessness,
stated linearly as
$
y_{xxx}\left(L\right)=0$ and $y_{xx}\left(L\right)=0,
$
respectively.
At the anchored end, we describe a polymer clamped to the
axoneme position $y_A(t)$, thus
$
y\left(0\right)=y_A(t);~~y_x\left(0\right)=0.
$
The above-listed boundary conditions constrain $4$
of the arbitrary constants, the remaining matching equations
constitute the remaining $4$, and we may thus completely specify the
solution.

\subsubsection{Construction of the inhomogeneous solution}
The governing equation is linear, allowing us
to separate $y$ into two separate solutions of
\rref{eqaf}: $y=y_0(h+g)$, where $y_0$ is some typical
length scale.
We choose $g$ ($h$) to
satisfy the (in)homogeneous boundary conditions and equation
of motion.
Compare this with the
first example solution of EHDI (section \ref{EHDIsec}),
in which $h$ was chosen to satisfy an inhomogeneous
{\it equation of motion} but homogeneous boundary conditions.

As described above, the axoneme moves with constant velocity
$v_c$ from $y_A= \Delta$ to $y_A= -\Delta$:
$
y_A\left(t\right)=-v_ct,\ \ (-{\Delta/ v_c}<t<{\Delta/ v_c}).
$
We must now merely solve for the inhomogeneous solution, expressed as
\beq
h^{\{a,f\}}\left(x,t\right)\equiv
\sum_{n=0}^{\infty}c_n^{\{a,f\}}\left(t\right)x^n.
\eeq
However, as in the derivation of \rref{EHDIP}, we expect
this solution to depend only linearly on the driving and
thus linearly on time. Constraining
$
\partial_t^2 c^{\{a,f\}}_n=0,
$
enforcing the 8 matching and boundary conditions,
and respecting the relationship between
$\p_{t}c_n$ and $c_{n+4}$
dictated by \rref{eqaf},
we
completely specify
the solution.
\subsubsection{Polynomial solution}
We here quote the solutions for $h\af$. In order to make
explicit the qualitative behavior
in limiting cases, and to write the solutions as
compactly as possible, we employ three nondimensionalized
variables:
\beq
\alpha\equiv {x\over x_p},~~
\sigma\equiv {{x-x_p}\over{L-x_p}}~,~~
 \tau\equiv  t{A\over\zeta x_p^4}.\label{NonDimVar}
\eeq
The coordinates $\{\alpha,\sigma\}\in(0,1)$
measure distance
on the left from the axoneme and on the right from the trap, while
$\tau$ is the
time rescaled by the characteristic elastohydrodynamic
time for the anchored section. The fact that $x_p$, rather
than $l\equiv L-x_p$, appears
explicitly in our choice of definition of
$\tau$ is reflected in the equation of motion, in that
the functions
now solve slightly different equations for the two sides:
\ba
h_{\tau}^a+{v_c\over y_0}{\zeta x_p^4\over
A}&=&-h^a_{\alpha\alpha\alpha\alpha};
\nonumber\\
h_{\tau}^f+
{v_c\over y_0}{\zeta x_p^4\over A}&=&-
\left({x_p\over l}\right)^4
h^f_{\sigma\sigma\sigma\sigma}.
\ea
The natural choice for $y_0$ is clearly
$y_0=v_c\zeta x_p^4/A$,
whereupon we are left with the dimensionless equations of motion
\ba
h_{\tau}^a+1&=-&h^a_{\alpha\alpha\alpha\alpha};~~\nonumber \\
h_{\tau}^f+1&=-&\rho^4h^f_{\sigma\sigma\sigma\sigma}~,
\ea
where we have introduced the ratio of lengths
\beq
\rho\equiv {x_p\over{L-x_p}}={x_p\over l},~~\label{rhodef}
\eeq
which describes the location of the point of
forcing $x_p$ relative to the ``extra" length $l$. As $x_p$ nears the
anchored end $x=0$, $\rho\goto 0$, and as $x_p$ nears the
free end $x=L$, $x_p/(L-x_p)=\rho\goto\infty$.

The expressions are then:
\ba
h^f\rho^5&=&-{\frac {1}{80}}\sigma^5
-{\frac {1}{24}}\rho\sigma^4
+\left({\frac {\rho}{6}}
+{1\over8}\right){\sigma}^{3}
-\frac {1}{4}\left (\rho+1\right ){\sigma}^{2}\nonumber\\
&&+\left ({\frac {1}{105}}\rho^4-{\frac {1}{8}}\rho^2-{\frac {1}{8}}\rho
+{\frac {3}{2}}\rho^4\tau\right )\sigma
\label{discrutch1}
\ea
\ba
h^a\rho^3&=&{\frac {1}{1680}}{\alpha}^{7}{\rho}^{3}
-{\frac {1}{240}}{\alpha}^{6}{\rho}^{3}-{\it \tau}\,{\rho}^{3}\nonumber\\
&&+\left ({\frac {13}{560}}\rho^3
-{\frac {1}{8}}\rho-{1\over8}
-{\frac {1}{2}}\rho^3\tau\right ){\alpha}^{3}\nonumber\\
&&+\left (-{\frac {11}{560}}\rho^3
+{\frac {1}{8}}\rho+{1\over8}
+{\frac {3}{2}}\rho^3\tau\right ){\alpha}^{2}
\label{discrutch2}.
\ea

\subsubsection{Decaying modes}

As described before, the solution $h$
to the inhomogeneous boundary conditions will describe
the long-time behavior remaining as transients decay exponentially.
We
now turn our attention to the transient $g$, which must
satisfy
the following
homogeneous boundary and matching conditions:
\ba
&g(0)=0,~~g_x(0)=0,\\
&g_{xx}(L)=0,~~g_{xxx}(L)=0\nonumber\\
&g\left(x_p\right)=0,~~
[g_x\left(x_p\right)]=0,~~
[g_{xx}\left(x_p\right)]=0,\nonumber
\ea
where the brackets indicate discontinuity.

The fact that all the boundary conditions are $0$-valued
means we stand a chance of
constructing
a self-adjoint operator, consistent with these conditions,
from the relevant differential
operator: $\partial_x^4$.
To do so, we left-multiply the
equation of motion by an as-yet arbitrary function $\Wk \left(x\right)$
and integrate over the {\it entire} domain. Since $h$ is constructed
to solve the (linear) equation of motion (\ref{eqndisc}),
$g$ must as well, and we derive an
equation of motion\footnote{The discontinuity in $y_{xxx}$ is
shared between $h$ and $g$, so we might for the sake of explicitness
include a term on the RHS of (\ref{gdyn}) proportional to $\W_k(x_p)$. However,
we will choose $\W_k(x_p)=0$ and thus this will not complicate
the behavior of $g_k(t)$.} for the quantities
$g_k\equiv \int_0^Ldx\W_k(x)g(x)$:
\beq
{1\over\nut}\partial_tg_k
=
-\int_0^L\!dx\W_k\left(x\right)g_{xxxx}
{}~.\label{gdyn}
\eeq
We then wish
to integrate the RHS by parts. However, we must admit the possibility
that $g$ supports a discontinuity in its higher-order
derivatives, \ie~ $[g_{xxx}\left(x_p\right)]\ne0$.
For this reason, we define $\{g^a\left(x,t\right),\Wk^a\left(x\right)\}$ and
$\{g^f\left(x,t\right),\Wk^f\left(x\right)\}$, as in \rref{eqaf},
defined on the anchored and free sections, respectively.
If we choose $\W$ to obey the same
homogeneous boundary
conditions as $g$, all surface terms from the integration by
parts in \rref{gdyn} vanish except
\ba
-\p_x\Wk^a\p_x^2g^a\left(x_p\right)&
+\p_x\Wk^f\p_x^2g^f\left({x_p}\right)&\nonumber\\
+\p_x^2\Wk^a\p_xg^a\left({x_p}\right)&
-\p_x^2\Wk^f\p_xg^f\left({x_p}\right)&
\ea
which vanish upon choosing $\W$ to obey the same
matching conditions as $g$.
The remaining equation is
\beq
-{1\over\nut}\p_tg_k=\int_0^{x_p}g^a\p_x^4\Wk^a
+\int_{x_p}^{L}g^f\p_x^4\Wk^f.
\eeq
We now choose $\Wk\af$ to obey the eigenvalue condition
$\p_x^4\Wk\af=k^4\Wk\af$. Since we have constructed
a self-adjoint operator, we can be sure the eigenvalues
are real and positive ({\cf} Appendix \ref{prop_app}). We are left with
\ba
-\p_tg_k
&=&\nut k^4\bigl\{\int_0^{x_p}\Wk^ag^a(x,t)
+\int_{x_p}^{L}\Wk^fg^f(x,t)\bigr\}\\
&=&\nut k^4g_k,
\ea
the solution of which is, as before,
$
g_k\left(t\right)=g_k\left(0\right){\rm e}^{-\nut k^4t}
$
for $k\ne0$. Fortunately, the boundary and matching
conditions do not admit a solution to $g_{xxxx}=0$,
and we need not consider a $0$ mode.

Given some initial condition $y(x,0)$, we project
it
onto this strange eigenspace spanned by $\{\Wk\}$.
We then construct $g\left(x,t\right)$
for all later times in a standard Green's function way,
\beq
g\left(x,t\right)=
\int_0^L\!\! {dx'\over L}
{\cal G}\left(x,x';t\right)\bigl
\{{y\left(x',0\right)\over y_0}-h\left(x',0\right)\bigr\}\label{gsolW}
\eeq
where
\beq
{\cal G}=\sum_{k}{\rm e}^{-\nut k^4 t}\Wk\left(x\right)\Wk\left(x'\right),
\eeq
and we see that all modes die.

\dofloatfig \begin{figure} \epsfxsize=3.0 truein
\centerline{\epsffile{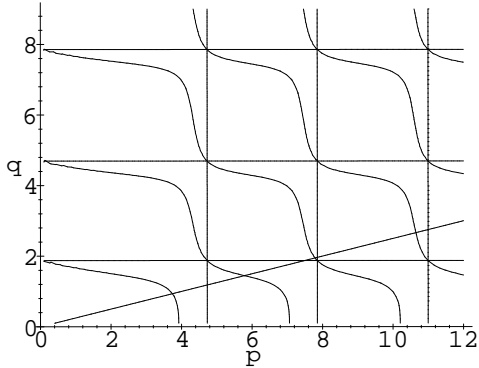}} \smallskip 
\caption[]{$p-q$ plane showing roots of the solvability condition. 
Here $p\equiv kx_p$ and $q\equiv kl=k(L-x_p)$. The geometry of the 
experiment dictates $\rho$, as described in the text. Vertical and 
horizontal lines correspond to solutions of $F(p)=0$ and $H(q)=0$, 
respectively. The diagonal line indicates the solution to $x_p/L=.8$.} 
\label{pqplot} \end{figure} \fi

\subsubsection{Construction of $\Wk$; solvability condition}

We now solve for the countably infinite sets $\{k\}$ and $\{\Wk\}$.
The general solution of $\p_x^4\Wk\af=k^4\Wk\af$ is
\ba
\Wk\af=
&a_1\af\sin\left(kx\right)+
&a_2\af\cos\left(kx\right)\nonumber\\
+&a_3\af\sinh\left(kx\right)+
&a_4\af\cosh\left(kx\right),
\ea
where the $8$ arbitrary constants will solve the $4$ boundary
and $4$ matching conditions. The insertion and elimination of
these constants is not a joyful task and will be omitted here.
The most important fact is that the set of $8$ equations
for $8$ unknowns can be written as a matrix of trigonometric
functions (as in Eq. \ref{mat})
annihilating
the vector ${\bf a}$ of unknowns.
The
$0$-valuedness
of the conditions dictates that the determinant ${\cal D}$ of this matrix
be $0$. This solvability condition is written explicitly in
appendix \ref{explicit_D};
the solutions, graphically constructed in Fig. \ref{pqplot}, determine
the allowed values of $k$ given some fixed ratio $\rho$.

Curiously,
the complicated solvability condition
can be expressed compactly as the separable equation
\beq
{\cal D}=\p_pF\left(p\right)H\left(q\right)
+F\left(p\right)\p_qH\left(q\right)=0\label{sepeqn},
\eeq
where $F(p)\equiv\cos(p)\cosh(p)-1$,
$H(q)\equiv\cos(q)\cosh(q)+1$, $p\equiv k x_p$,
and $q\equiv kl$.
This differential relation describes the motion along
each of the branches shown in the figure as $\rho$ varies,
each branch indexed
by arbitrary constants introduced upon integrations of \rref{sepeqn},
and separated by the singularity lines $\{H=0,F=0\}$ at which the
differential equation is not invertible.
The geometry chosen by the experimenter dictates $x_p, l$, and
therefore $x_p/l={p/ q}=\rho$. Inspecting the figure, we see
that we choose a set of modes by drawing a line of slope $\rho$
through the origin; each intersection with a branch corresponds to
one mode.

A further curiosity is that the each of the four
equations $\{F,H,F_p,H_q\}=0$ is itself a separate solvability condition
associated with a separate experimental geometry. In the
language of Appendix \ref{explicit_D}, we may rewrite the
condition ${\cal D}=0$, with the first
letter below indicating the boundary condition for the left end of one
side, and the second for the condition at the right,
\beq
{\cal D}=(cf)(hf)+(ch)(ff).
\eeq

\dofloatfig \begin{figure} \epsfxsize=2.8 truein
\centerline{\epsffile{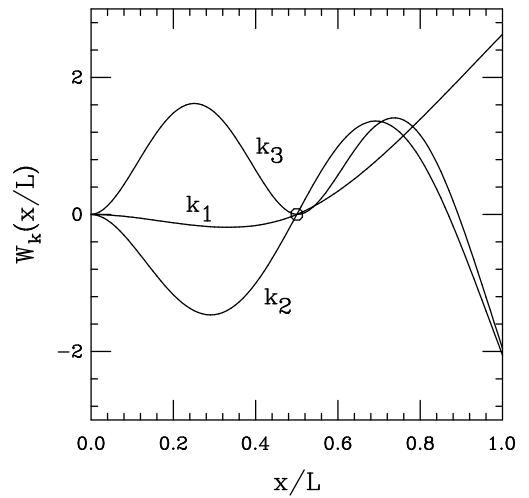}} \smallskip 
\caption[]{The first three modes for the discontinuous EHDI experiment.  
The circle indicates the pinning point $x_p/L=p/(p+q)$.} 
\label{024plot} \end{figure} \fi

Otherwise stated, the total solvability condition is
an average of those for (i) the left side clamped at $0$ and
free at $x_p$, with the right side hinged at $x_p$
and free at $L$, and (ii) the left side clamped at $0$ and
hinged at $x_p$, with the right side free at $x_p$ and
free at $L$.

\subsubsection{Initial data}

Armed with a set of decaying modes, we may solve the equation for
all times.
We now attempt, as before, to construct a polynomial solution which
describes the initial data. The polymer sits
at rest with $y(0)=\Delta$ and is subject to the stated matching
and boundary conditions.
Noting that the shape must be described by
polynomials of less than fourth order, since we wish to
describe an elastica experiencing no forces,
we find for the new polynomial $\bar y$
\ba
{\bar
y}^a\left(\alpha\right)&=&\Delta\left(1-{3\over2}\alpha^2+\oh\alpha^3\right)\\
{\bar y}^f\left(\sigma\right)
&=&-\Delta{\frac {3\,\sigma}{2\,\rho}}~.\label{yinff}
\ea
The final configuration, after the axoneme comes to rest and
all transients have died, will be $-\bar y$: a third-order
polynomial on the left and a straight line on the right. A pleasant
fact is that this polynomial can also be derived by taking
the limit as $v_c\goto 0$ of $y_0h^{\{a,f\}}$.

\subsubsection{Comparison with experiment}

We now wish to use this information
to
arrive at a measurement of $A$. Hoping to
verify the plausibility of this analysis, we
compare with the results and accompanying model
published with the experiment.

The model presented in \cite{felg} can be summarized
as follows.
First, the forces
are calculated on a slender body subject to (i) a constant
flow and (ii) a flow linearly increasing from $0$ at the origin
to some $v_m$ at the end.
The linearly-increasing
flow describes that experienced
by a rigidly rotating rod, in hopes that the
force experienced by the actual (curved) filament,
itself neither a straight rod, nor moving with constant velocity in time,
is well-approximated.
Curiously, different drag coefficients are used for these two forces.

The forces are then used to compute displacements, using
the results for the elastica clamped at the origin,
a condition which is unfortunately consistent with neither
the assumption of
a rodlike shape
nor
with the experimental
geometry, in which the free side of the elastica is
neither clamped ($y_x=0$, as was assumed for the calculation of displacement)
nor hinged $(y_{xx}=0$, as might be considered appropriate
for a straight rod); it is in
fact experiencing a torque at $x_p$ due to the
side of the filament between the trap and the axoneme.

Summing (i) the displacement $y_c(v_c;\zeta_1)$ of a clamped elastica
in constant flow (using the first drag coefficient) and (ii) the
displacement $y_l(v_m;\zeta_2)$ of an elastica
subject to the linearly increasing flow it would have experienced were it a
rigid rod
(using the second drag coefficient),
the total deflection $y_{\delta}$
from a straight line is obtained (see Eq. 20 of \cite{felg}):
\beq
y_{\delta}={\pi\eta(11 v_m+30 v_c)l^4\over 60 A\ln(2l/d)}.\label{ydelta}
\eeq

It is then assumed that this displacement equals the deflection
which would have been experienced had the filament
not been initially horizontal,
but rather constrained to some nonzero slope at the trap,
specifically, that of the $t\goto\infty$ solution for the shape, $-{\bar y}$.

This deflection $y_{\delta}$,
from the
right tip position
at $t=\infty$
to that at $t=\Delta/v_c$,
when the axoneme halts, is the first experimental observable. The
second is the velocity at this tip at $t=\Delta/v_c$,
which is then equated with the  maximum
velocity $v_m$ in the expression for $y_l$.
Dividing $y_{\delta}$ by the weighted sum of velocities
appearing in the numerator, one obtains the combination
of two observables ($y_{\delta}, v_m$) and one experimental parameter
($v_c$) which, according to this model, is equivalent
to a simple quotient with units of time and in which $A$ appears
explicitly,
\beq
{\cal T}={y_{\delta}\over (11 v_m+30 v_c)}={11\pi\eta\l^4\over \Afelg
60\ln(2l/d)}
\eeq
where $\Afelg$ indicates the rigidity which would
be extracted from the data using this model. Note that the expression
is a function only of $l$, the length of the free segment of
the polymer. In this model, $l$ is taken to be ``the
hydrodynamically relevant length"\cite{felg}.

Returning to the PDE-treatment of the problem, we see that
the solutions to \rref{eqaf} for the long-time polynomial shape
relate
the velocity of the free end
to the constant stage velocity as
\beq
v_m={v_c}h^f_{\tau}\vert_{\sigma=1}={3\over 2}{l \over x_p}v_c\label{v_relatn}.
\eeq
This simplification allows us to solve for the
ratio in (\ref{ydelta})
of observables and parameters in terms of the polynomials in Eqs.
\ref{yinff} and \ref{discrutch1}.
\beq
{\cal T}=
{\zeta l^4\over2520 \Aus}
{\frac
	{\left (-16\,{\rho}^{4}+210\,{\rho}^{2}+420\,\rho+231\right )}
{\left(11+20\,\rho\right)}}
\eeq
where $\Aus$ indicates the value of the bending
modulus which one would calculate had one used the same
data and thus the same quotient ${\cal T}$.

\begin{figure} \epsfxsize=3.3 truein
\centerline{\epsffile{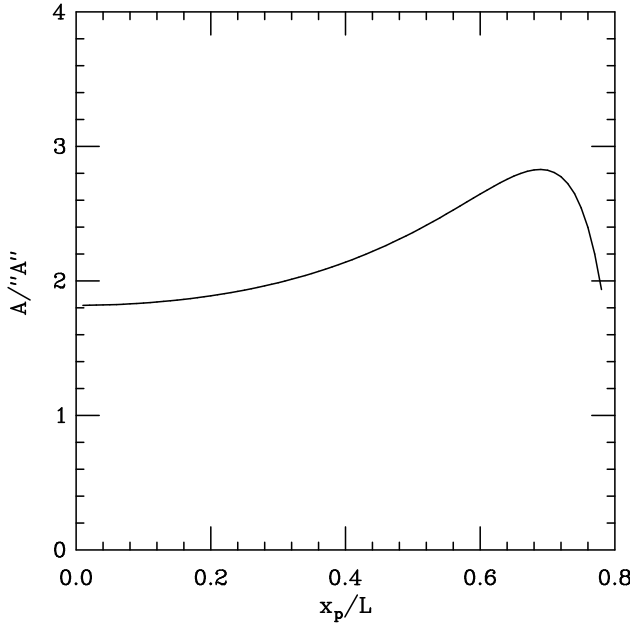}} \smallskip 
\caption[]{Quotient of values derived for the bending modulus 
using the model of \cite{felg} and that presented herein.} 
\label{err_fig} \end{figure}

The ratio $\Aus/\Afelg$ is a simple expression
and allows us to compare by what factor this differential equation-based
analysis
differs from the published model given some experimental
data ${\cal T}(l)$.
Equating the two expressions for ${\cal T}$ and isolating
$\Aus/\Afelg$, we find

\beq
{
\Aus
\over
\Afelg
}
=
2
{
\ln({L\over d}{1\over 2(1+\rho)})
\over
\ln({L\over d}{4\over \sqrt{e}})
}
\left(1+
{
{210\rho^2-16\rho^4}
\over
{231+420\rho}
}
\right)
\eeq
where we have moved the ${\cal O}(1)$ constants into
the argument of the logarithm for compactness.
The dominant behavior is captured by the polynomial dependence
on $\rho$, but we can see that the function will
be $\sim2$, indicating a systematic underestimate
by the published model of the bending modulus.
A plot of the ratio, with typical
values taken from experiment, is shown in Fig. \ref{err_fig}.

\dofloatfig  \fi

Note that we have arrived at this expression by ignoring the
transient component in the exact expression. This is valid only
in the range where the polynomial (asymptotic) solution
dominates over the transients, a condition which is violated
above $\rho\simeq4, x_p/L=\rho/(\rho+1)\simeq 0.8$.

We expect then, for results obtained with the model in
\cite{felg} systematically to underestimate the bending
modulus $A$ by a factor $\sim1/2$. In fact, we can fit the
published values of ${\cal T}(l)$ via nonlinear $\chi^2$ fitting
to arrive at a value for A. Unfortunately, the incorporation of
the effects due to the left end of the polymer introduces
a dependence on $L$, data for which are unpublished.
Unable to elicit a response from the authors
regarding the total lengths
used in the experiment,
we could only fit for both parameters.
The analysis produced lower $\chi^2$ values than the model of
\cite{felg}; however, $\chi^2$ is extremely weakly-dependent
on $L$ and not a particularly advantageous method without this datum.
In each case, the fit values of $A$ were approximately
twice that reported in \cite{felg}.

\bigskip
\noindent

\section{Conclusions}

We have attempted to show that a systematic treatment of linearized
elastohydrodynamics for filamentous biopolymers can be formulated
with fruitful results.
Specifically, we anticipate these methods should be
useful in the design and analysis of dynamic experiments.
Further, we have proposed a new technique (EHDII) which
exploits viscous hydrodynamics to extend the range
of mechanical experiments of bending moduli
to more flexible polymers.
We expect this experiment to produce more
accurate results when repeated with lower-amplitude driving, and thereby
help determine conclusively the existence or
nonexistence of scale-dependent or time-dependent elastic behaviors
in biopolymers as well as the value of $A$ in {in vitro} assays.
It is our hope that the analysis associated with this experiment will also
encourage renewed interest in the problems of flagellar
motion and slender-body hydrodynamics in general.

Moreover, we have seen that attentiveness to
equations of motion and boundary conditions for the elastica has
measurable consequences, and that
construction of the
appropriate function space associated with these conditions
leads to a pleasant union of mathematical and physical consequences,
quantifying our intuitions and relating transcendental equations to
physical effects and experimental observables.
We believe that the significance of boundary conditions
and the natural function space for the elastica has been
overlooked in existing treatments of the dynamics
and statistical mechanics relevant to experiments being
performed and discussed by the community.

We look forward to the extension of this analysis
to arbitrary geometries, reflecting distortions
beyond small order and hydrodynamics beyond the lowest-order
approximation of slender-body flow. Though the exact
equation of motion is nonlinear, the Stokesian dynamic
and the vanishing of successive derivatives remain, and
we expect the mechanisms and effects which we have
outlined here to persist.

Natural
extensions of this research include nonplanar geometries and
the incorporation of twist.
These would be complementary to recent work on the {\it Hamiltonian}
dynamics of twisted elastic rods \cite{tabor,wilma}.
The dynamics of twist are especially
intriguing in the light of recent work on twist-bend
coupling \cite{randy}
and the proposal that this coupling creates scale-dependent
elasticity in actin \cite{Kas}.
Another intriguing experimental
realization of nonplanar viscous elastohydrodynamics with twist
is the supercoiling of fibers of mutants of the common bacterium
{\it Bacillus subtilis} \cite{neil}. We are currently
formulating the analysis appropriate to these promising
applications,
armed with the lessons learned in this investigation.
\section{Acknowledgments}
We thank Steve Block and Steve Gross for early discussions
on analysis of the data and for bringing \cite{felg} to
our attention, and Joseph K\"as for pointing out \cite{Gittes}.
We also thank Almut Bruchard,
Mark Johnson,
Frank J\"ulicher,
Randall Kamien,
David Levermore,
Mike Shelley, and
especially Tom Powers
for useful conversations and insight.
This work was supported by NSF PFF Grant DMR 93-50227 and the
A.P. Sloan Foundation (REG).

\appendix
\section{Eigenfunctions of $\partial_{\alpha}^4$}
\label{family}
If we equate functions related by the  reflection
$\alpha\goto 1-\alpha$,
there are $4(4+1)/2=10$ distinct eigenfunctions of $\partial_{\alpha}^4$,
determined by boundary conditions,
for which this
operator is self-adjoint. Each has an associated solvability condition
for the eigenvalues $k$. We list the solvability conditions and the
unnormalized eigenfunctions, indexed  according to the conditions at
the ends:
\ba
\left(f\right)&{\rm free}:&h_{xx}=h_{xxx}=0\\
\left(c\right)&~{\rm clamped}:~&h=h_{x}=0\\
\left(h\right)&{\rm hinged}:&h=h_{xx}=0\\
\left(t\right)&~{\rm torqued}:~&h_{xxx}=h_x=0
\ea
The general solution is $\PHI_k=a_1\cos k \alpha + a_2\sin k \alpha
+a_3 \cosh k \alpha + a_4 \sinh k \alpha $, where $\alpha\in(0,1)$.
The letters {f,c,h,t}  denote
the boundary conditions at the left and right. Note that
for special cases, the calculated $\PHI_k$ are merely a Fourier
basis.
\ba
\bullet{\rm f-f}:&& \\\label{A5}
&&\cos k\cosh k=1; \nonumber \\
&&\PHI_k=
\left(\sin k\alpha+\sinh k\alpha\right)\left(\sin k+\sinh k\right)\nonumber \\
&&~+
\left(\cos k \alpha + \cosh k \alpha \right)
\left(\cos k-\cosh k\right)\nonumber
\nonumber\\
\bullet{\rm c-c}:&& \\
&&\cos k\cosh k=1;\nonumber \\
&&\PHI_k=
\left(\sin k\alpha-\sinh k\alpha\right)\left(\sin k+\sinh k\right)\nonumber \\
&&~+
\left(+\cos k \alpha - \cosh k \alpha \right)
\left(\cos k-\cosh k\right)
\nonumber\\
\bullet{\rm f-c}:&&\\
&&\cos k \cosh k = -1;\nonumber \\
&&\PHI_k=
\left(\sin k\alpha+\sinh k\alpha\right)\left(\sin k-\sinh k\right)\nonumber \\
&&~+
\left(\cos k \alpha + \cosh k \alpha \right)
\left(\cos k+\cosh k\right)
\nonumber\\
\bullet{\rm h-h}:&& \\
&&\sin k=0;\nonumber \\
&&\PHI_k=\sin\ka \nonumber\\
\bullet{\rm t-t}:&& \\
&&\sin k=0;\nonumber\\
&&\PHI_k=\cos\ka \nonumber\\
\bullet{\rm h-t}:&&  \\
&&\cos k =0;\nonumber \\
&&\PHI_k=
\sin\ka \nonumber\\
\bullet{\rm f-h}:&& \\
&&\tan k=\tanh k;\nonumber \\
&&\PHI_k=
\left( \sin k\alpha+\sinh k\alpha\right)
\left(-\cos k+\cosh k\right)\nonumber \\
&&~+
\left(\cos k \alpha + \cosh k \alpha \right)
\left(\sin k-\sinh k\right)
\nonumber\\
\bullet{\rm f-t}:&& \\
&&\tan k = -\tanh k;\nonumber \\
&&\PHI_k=
\left(\sin k\alpha+\sinh k\alpha\right)\left(\sin k+\sinh k\right)\nonumber \\
&&~+
\left(\cos k \alpha + \cosh k \alpha \right)
\left(\cos k-\cosh k\right)
\nonumber\\
\bullet{\rm c-t}:&&\\
&&\tanh  k = -\tanh k;\nonumber \\
&&\PHI_k=
\left( \sin k\alpha-\sinh k\alpha\right)\left(\sin k+\sinh k\right)\nonumber \\
&&~+
\left( \cos k \alpha - \cosh k \alpha \right)
\left(\cos k-\cosh k\right)
 \nonumber\\
\bullet{\rm c-h}:&&  \\
&&\tan k = \tanh k;\nonumber \\
&&\PHI_k=
\left(\sin k\alpha-\sinh k\alpha\right)\left(\cos k+\cosh k\right)\nonumber \\
&&~+
\left(-\cos k \alpha + \cosh k \alpha \right)
\left(\sin k+\sinh k\right)\nonumber
\ea

\section{Properties of the $\PHI_{\rm k}$ basis}
\label{prop_app}

We define $\PHI_k$ by the operator
of which it is an eigenfunction and by its boundary conditions:
\beq
{\cal H}\equiv \partial_{\alpha}^4~ + {\rm boundary~conditions}.
\eeq
Next we establish the self-adjointness of ${\cal H}$, which we
may express in
Dirac notation as
$\langle\PHI_k|{\cal H}\PHI_j\rangle=
\langle{\cal H}\PHI_k|\PHI_j\rangle$.
This follows from the integration
\ba
\int_0^1\! d\alpha \PHI_k\partial_{\alpha}^4\left(\PHI_j\right)&=&
\int_0^1\!d{\alpha}\partial_{\alpha}^4\left(\PHI_k\right)\PHI_j \\
&&+\PHI_k\da{3}\PHI_j\bigl|_0^1-\da{1}\PHI_k\da{2}\PHI_j\bigl|_0^1\nonumber \\
&&+\da{2}\PHI_k\da{1}\PHI_j\bigl|_0^1
-\da{3}\PHI_k\PHI_j\bigl|_0^1\nonumber\\
&=&\int_0^1\!d{\alpha}\partial_{\alpha}^4\left(\PHI_k\right)\PHI_j~,\nonumber\
\ \ \ \Box
\ea
where all the surface terms are seen to disappear
for any choice of boundary conditions from appendix \ref{family}.

We can also establish the orthogonality of the members
of a family since
\ba
\int_0^1d\alpha\PHI_j\PHI_k&=&\int_0^1d\alpha\PHI_j{\da{4}\PHI_k\over k^4}=
k^{-4}\int_0^1d\alpha\da{4}\PHI_j\PHI_k \nonumber \\
&=&\Bigl({j\over k}\Bigr)^{4}\int_0^1d\alpha\PHI_j\PHI_k~.
\ea
Thus $\int_0^1d\alpha\PHI_j\PHI_k=0$ for $j\ne k$.

We justify considering only real, positive values of
$k^4$ by observing
\ba
k^4\int_0^1d\alpha\PHI_k\PHI_k^{*}
&=&\int_0^1d\alpha\da{4}\PHI_k\PHI_k^{*}\nonumber \\
&=&\da{3}\PHI_k\PHI_k^{*}\vert_0^1d\alpha-\da{2}\PHI_k\da{1}\PHI_k^{*}\vert_0^1d\alpha
\nonumber \\
&&+\int_0^1d\alpha\da{2}\PHI_k\da{2}\PHI_k^{*}
\ea
so
\beq
k^4={\int_0^1d\alpha|\da{2}\PHI_k|^2\over\int_0^1d\alpha|\PHI_k|^2}~,
\eeq
the quotient of two real, positive quantities,
which is necessarily real and positive.

\section{Exact solution to EHD problem II}
\label{exactsolnapp}

The exact solution to EHDII can be written in somewhat compact form
at the cost of introducing new definitions.
Employing the parameter $\alpha\equiv x/L$, the
rescaled length
${\cal L}\equiv L/\ell$, and
the constant $z_0=exp(-i\pi/8)$ as in \rref{z0def}, we introduce
\beq
\xi \equiv  {\rm e}^{z_0{\cal L}}
\eeq
and then write
\beq
h(\alpha,{\cal L})\equiv h_\infty(\alpha)+h_{\flat}(\alpha,{\cal L})~.
\eeq
where $h_\infty$ is the semiinfinite solution, and
$h_{\flat}\goto 0$ as ${\cal L} \to \infty$.
Explicitly,
\beq
h_{\flat}={1\over 2}C_1\left(\xi^{\alpha}-\xi^{-\alpha}\right)
+{1\over 2}C_2\left(\xi^{i\alpha}-\xi^{-i\alpha}\right)~,
\eeq
where
\ba
C_1({\cal L})&=&\xi^{-1}{(-i\xi^{-i}-\xi^{i}+(1+i)\xi)\over
(i\xi^{-1-i}+\xi^{-1+i}-\xi^{1-i}- i\xi^{1+i})}\nonumber \\
C_2({\cal L})&=&\xi^{-i}{(i\xi^{-1}-\xi^{1}+(1-i)\xi^{i})
\over (-i\xi^{-1-i}-\xi^{-1+i}+\xi^{1-i}+i\xi
^{1+i})}~,
\ea
and
\beq
h_\infty=\oh\left(\xi^{-\alpha}+\xi^{-i\alpha}\right).
\eeq

\section{Solvability condition for discontinuous EHDI}

\label{explicit_D}

Associated with the discontinuous EHDI problem is the set of
eigenvalues of $\da{4}$ appropriate to the boundary and matching
conditions. Like the statement $\sin(k L)=0$ for determining the
allowed Fourier $k$ values for a doubly-hinged filament of length $L$, there is
a solvability condition
for
$p\equiv  k x_p$
,
$q\equiv k \left(L-x_p\right)=kl$,
derived by setting the determinant of
an $8$ by $8$ matrix to $0$, an equation
which can
be written:
\ba
0&=&{\cal D}\left(p,q\right)=\nonumber \\
&&-\sin\left(q\right)\cos\left(p\right)\cosh\left(q\right)\cosh\left(p\right)
\nonumber \\
&&-\cos\left(q\right)\sin\left(p\right)\cosh\left(q\right)\cosh\left(p\right)
\nonumber \\
&& +\cos\left(q\right)\cos\left(p\right)\sinh\left(q\right)\cosh\left(p\right)
\nonumber \\
&&+\cos\left(q\right)\cos\left(p\right)\cosh\left(q\right)\sinh\left(p\right)
\nonumber \\
&&-\sin\left(q\right)\cosh\left(q\right) +\cos\left(q\right)\sinh\left(q\right)
\nonumber \\
&& +\sin\left(p\right)\cosh\left(p\right)
-\cos\left(p\right)\sinh\left(p\right)
\ea
The ratio $x_p/(L-x_p)=q/p=\rho$, set by the geometry of the
experiment, fixes a diagonal line passing through
the origin and intersecting the set of curves defined by this equation
to define all the allowed $k$ values, as illustrated
in Fig. \ref{pqplot}.

\end{document}